\documentclass[final,5p,times]{elsarticle}

\usepackage{amssymb}
\usepackage{amsmath}
\graphicspath{ {images/} }

\journal{Astronomy and Computing}

\begin{document}

\begin{frontmatter}



\title{Transient Simulations for Radio Surveys}


\author[GWU,APSIS]{Sarah I Chastain\corref{cor1}}

\affiliation[GWU]{organization={Department of Physics, The George Washington University},
            city={Washington},
            postcode={20052}, 
            state={DC},
            country={USA}}
\cortext[cor1]{Corresponding author}
\affiliation[APSIS]{organization={Astronomy, Physics and Statistics Institute of Sciences (APSIS), The George Washington University},
            city={Washington},
            postcode={20052}, 
            state={DC},
            country={USA}}
 \affiliation[UVI]{organization={University of the Virgin Islands},
 	city={Charlotte Amalie},
    postcode={00802},
    state={USVI},
    country={USA}}

\author[GWU,APSIS]{Alexander J van der Horst}
\author[UVI]{Dario Carbone}


\begin{abstract}
Several new radio facilities have a field of view and sensitivity well suited for transient searches. This makes it more important than ever to accurately determine transient rates in radio surveys. The work presented here seeks to do this task by using Monte-Carlo simulations. In particular, the user inputs either a real or simulated observational setup, and the simulations code calculates transient rate as a function of transient duration and peak flux. These simulations allow for simulating a wide variety of scenarios including observations with varying sensitivities and durations, multiple overlapping telescope pointings, and a wide variety of light curve shapes with the user having the ability to easily add more. While the current scientific focus is on the radio regime, with examples given here from the MeerKAT telescope in South Africa, the simulations code can be easily adapted to other wavelength regimes.
\end{abstract}




\end{frontmatter}


\section{Introduction}
\label{sec:sample1}

We are entering an exciting era in time-domain astronomy.  New and upgraded facilities such as the Vera C. Rubin Observatory \cite{2019ApJ...873..111I} and Zwicky Transient Facility \cite{2019PASP..131a8002B} in the optical, and the MeerKAT \cite{2016mks..confE...1J} and Australian Square Kilometer Array Pathfiner (ASKAP) \cite{2021PASA...38...54M} radio telescopes, have been finding, or are expected to find, transients and variables in images at rates that are orders of magnitude higher than ever before. This is in addition to exciting new transients found in time series data, such as the wealth of Fast Radio Bursts (FRBs) found using the Canadian Hydrogen Intensity Mapping Experiment (CHIME) \cite{2019Natur.566..235C}. 

Many transients are discovered in blind searches, found by examining large portions of the sky for new sources or known sources that display significant flux changes. There are also transients found in a targeted way, such as those associated with gravitational wave events \cite{2017PhRvL.119p1101A,2017Natur.551...71T,2018ApJ...868L..11M}, gamma ray bursts \cite{1997Natur.389..261F,2018MNRAS.473.1512A}, tidal disruption events \cite{2011Sci...333..199L,2016Sci...351...62V}, and outbursts from X-ray binaries \cite{2004MNRAS.355.1105F,2017MNRAS.469.3141T}. Considering that transients can be found in both blind and targeted searches brings up important questions: if we are doing a targeted transient search, what is the chance that a detection may be a different transient source that happens to be in the same area of the sky, even within the same uncertainty region of the transient of interest? How many transients of a certain type or with a specific light curve shape would we expect to find in a given survey? Finding the answers to these questions is important for a variety of applications in time-domain astronomy and requires calculating transient rates with high accuracy. 

The most straightforward approach to calculating a transient rate is to use the Poisson distribution to find a rate given the number of detections in a survey, but there are shortcomings in this simplified approach \cite{2016MNRAS.459.3161C}. This transient rate does not account for a number of important factors such as the relative timescales of the transients and the observations, and some of the confounding observational effects such as gaps within an observation or a survey. In addition, it does not account for the distribution of sensitivities present in the observations of a real survey. These effects can be partly mitigated in an analytical approach \cite{2016MNRAS.459.3161C}, but Monte-Carlo simulations provide a way to more easily account for the issues presented by real observations and surveys in transient rate calculations \cite{2017MNRAS.465.4106C}. 

\citet{2017MNRAS.465.4106C} examined two light curve shapes: the tophat light curve, a light curve that instantaneously rises to its peak flux and at some point in time instantaneously decays; and the fast rise exponential decay (FRED) light curve, a light curve that instantaneously rises to its peak flux and exponentially decays thereafter. The differences between the resulting transient rates from these two light curve shapes indicate how the wide variety of real light curves can affect transient rate calculations. This is in addition to the previously mentioned observational effects that should be accounted for. 

Observational radio surveys present a number of challenges for computing transient rates. Although radio observations can be calibrated using a sky model, many radio observations require the use of calibrator fields that need to be observed at certain time intervals before, after, and during an observation of a science target field. This means that the telescope does not continuously point at a target for an entire observation. Often when using a calibrator source, a radio observation would be broken down into observing a very bright, well-known source to calibrate the bandpass, followed by alternately observing a bright source close to the target for gain calibration and the science target. This means that the time on target is less than the total observing time, and that there are gaps in the target observations. This also means that there is the possibility of searching calibrator fields for transients \citep{2011ApJ...728L..14B}. Furthermore, typical radio observations can be broken down into shorter timescales for imaging. In addition, in order to explore a wider field of view, a survey may consist of multiple adjacent pointings on the sky with some degree of overlap between pointings. These pointings may have different limits on transient rates due to differing observing cadences, and the overlap regions will provide different transient rates as well.

The goal of this work is to calculate transient rates while accounting for the aforementioned features and complexities of radio surveys. Corrections for observational effects such as gaps in observations, systematic errors in flux measurements, different kinds of transient light curves, multiple overlapping pointings, and a distribution of observational sensitivities are accounted for. Mitigating all the aforementioned effects makes the transient rate calculations more accurate. In addition, the publicly available simulations code is relatively simple in its use, has Python 3 support, and is designed for modularity so that the user can easily add new items such as other light curve shapes than those already provided.

In section~\ref{design}, we will go into detail on how the code is written and its features implemented. We will in section~\ref{sec:results} present and discuss results from several example radio surveys illustrating the various features. In section~\ref{performance} we will look at the computational performance of the simulations code. In section~\ref{futureapplications} we will discuss the ways in which this code can be expanded in the future, and we draw conclusions in section~\ref{conclusions}.

\section{Design} \label{design}

\subsection{Language and Libraries}
The code was written in Python\footnote{http://www.python.org} and designed for its most up-to-date versions ($>3.6$). It uses several libraries: Astropy \cite{2013A&A...558A..33A} to provide accurate angular source separation calculations and any necessary coordinate system changes; Scipy \cite{2020NatMe..17..261V} for a few special mathematical functions; Bitarray\footnote{https://github.com/ilanschnell/bitarray} for storing large amounts of information efficiently; tqdm\footnote{https://tqdm.github.io} for easy-to-use progress bars; and Numpy\footnote{https://numpy.org/} for the vast majority of the numerical computations. In addition, the script to assist with creating input files uses Common Astronomy Software Applications (CASA) \cite{2007ASPC..376..127M} to read and extract metadata from radio measurement sets. 

By using an interpreted language that allows for the use of classes, the code is easy to modify or extend for different use cases, or to increase accuracy. Adding new light curves can be done by creating a Python file with the name of the light curve and a class with the essential information. Using Numpy partially makes up for Python's lack of speed compared to a compiled language such as C. The information on whether a simulated source is detected is stored and written using bit arrays, in order to reduce memory usage so that these computations can be performed on a regular desktop or laptop. 

\subsection{Input}
In order to accurately simulate transient rates, it is necessary to provide detailed information on the survey that will be simulated. This information includes observation times, pointings, field of view, sensitivity, and any gaps in the observations. This information is either supplied by the user as a comma-separated values (CSV) file or it can be generated using a separate script that extracts information from the metadata in the measurement sets of the survey observations.

In addition, the simulations code base contains a configuration file with settings that can be adjusted depending on the use case. These settings include items such as number of transients to simulate, number of transient detections in the survey, flux and duration ranges to simulate, detection threshold, light curve type, confidence level, output filename, and options for simulating a survey such as number of observations, sensitivity, mean and standard deviation of simulated normally-distributed error in sensitivity, interval between observations, and the duration of the observations. The light curve type can be any of the included ones or a new light curve created by the user. 

\begin{figure}[h]
\includegraphics[width=\columnwidth]{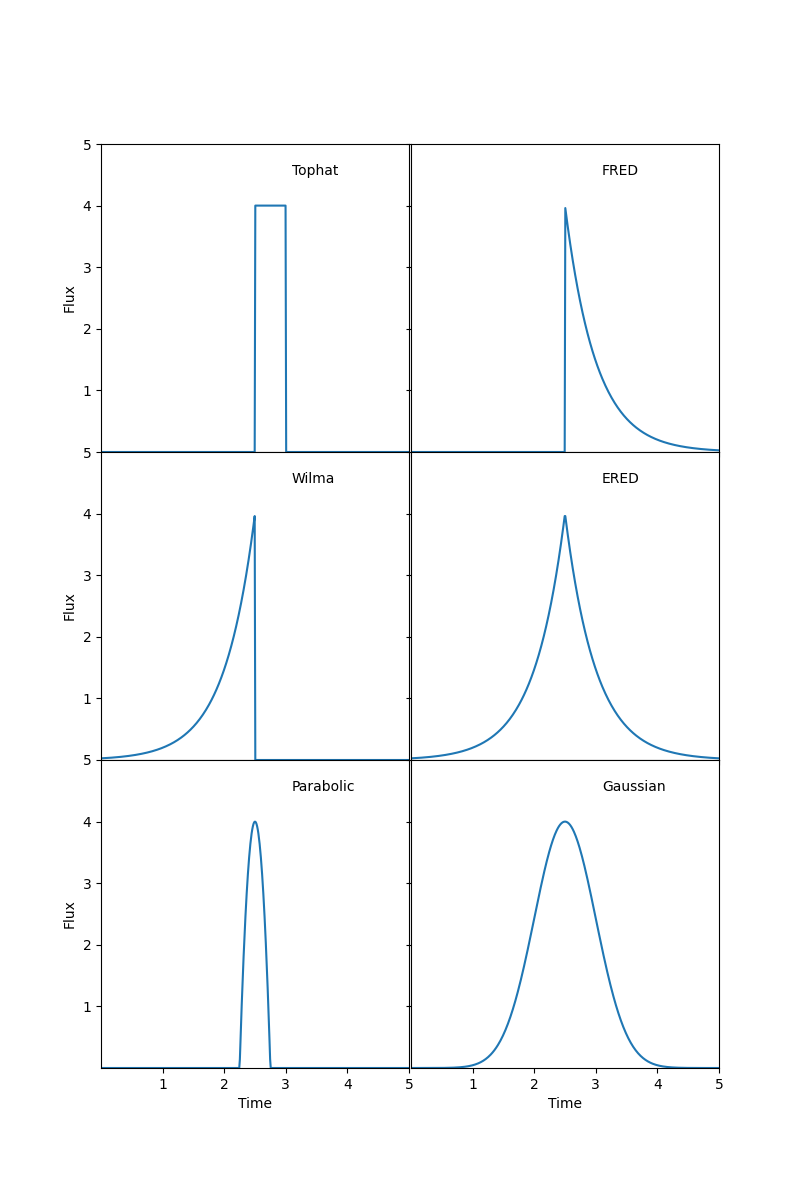}
\caption{Examples of all light curve shapes included in the simulations in this paper, plotted using arbitrary units of time and flux.}
\label{multilc}
 \end{figure}
\subsection{Light Curves}
In Figure~\ref{multilc} we show the light curves included in the simulations, which are the tophat, fast rise exponential decay (FRED), exponential rise fast decay (Wilma), exponential rise exponential decay (ERED), Gaussian, and parabola. The tophat light curve is defined to have an instantaneous rise to the peak flux, followed at some point in time by an instantaneous decay. This light curve represents the classic case of a transient that turns on and off, and the simplest form of transient light curves. The FRED light curve instantaneously rises to the peak flux and exponentially decays. This light curve is commonly observed in a variety of X-ray and gamma-ray transients. The Wilma is simply the time reversed FRED: it exponentially rises to the peak flux and then instantaneously decays. Including the Wilma light curve is a convenient way to introduce the simplest form of light curve with no definite start or end. The ERED is such a light curve that is formed by putting the previous two together: it exponentially rises to a peak flux and then exponentially decays. The Gaussian is a light curve that has the shape of a Gaussian function with the mean being located at the peak flux and the duration given by the standard deviation. Similar light curves can be seen arising from, for example, binary systems and magnetar bursts.  The parabolic light curve is a concave down parabola that reaches the peak flux at the vertex and the duration being the range of time in which the flux is positive; its inclusion provides an example of a light curve with a definite duration but with a profile that rises and falls below the peak flux in a symmetric way (i.e., one step more complex than the tophat). More details about these light curves, including their mathematical definitions, can be found in~\ref{sec:lc:appendix}.

\begin{figure}[h]
\includegraphics[scale=0.5]{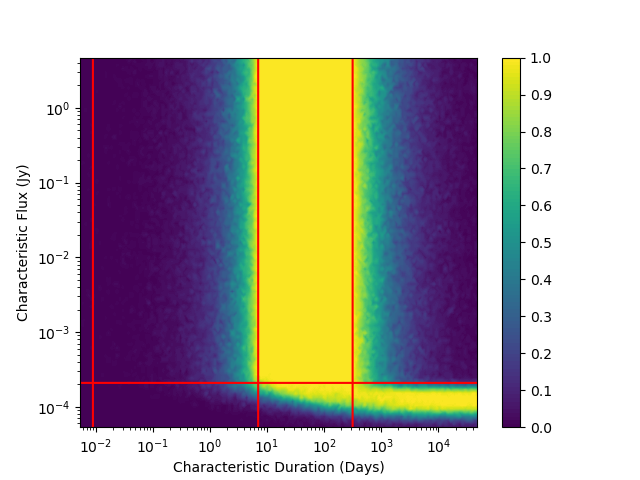}
\caption{Probability contours for the tophat light curve. The leftmost vertical line marks the shortest observation in the survey, the middle line corresponds to length of the longest gap between observations, and the rightmost line corresponds to approximately the length of the survey itself. The horizontal line is a line marking the flux value that is greater than 99\% of the flux values of the false transient sources.}
\label{tophat}
 \end{figure}

\begin{figure}[h]
\includegraphics[scale=0.5]{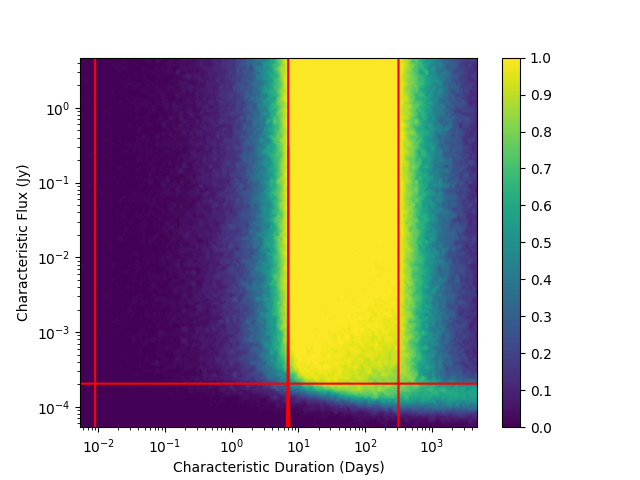}
\caption{Probability contours for the parabolic light curve. The meaning of the horizontal and vertical lines is the same as in Figure~\ref{tophat}.}
\label{parabolic}
 \end{figure}

\begin{figure}[h]
\includegraphics[scale=0.5]{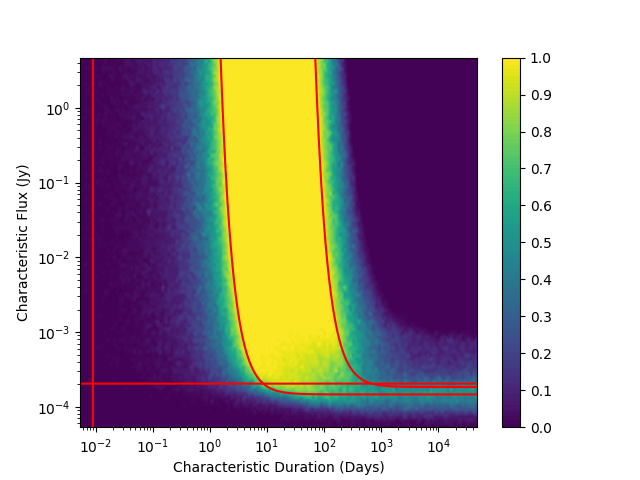}
\caption{Probability contours for the Gaussian light curve. The leftmost vertical line marks the shortest observation in the survey, the curve in the middle marks a boundary in the duration of sources to the left of which these sources can fall in the longest gap between observations and go undetected in any observations, and the rightmost curve is a boundary to the right of which sources can be detected as a constant source by being detected in every observation. The meaning of the horizontal line is the same as in Figure~\ref{tophat}.}
\label{gaussian}
 \end{figure}

\begin{figure}[h]
\includegraphics[scale=0.5]{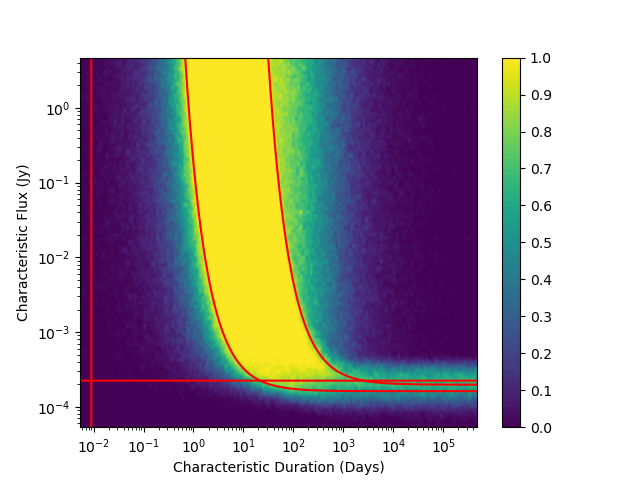}
\caption{Probability contours for the FRED light curve. The meaning of the horizontal and vertical lines is the same as in Figure~\ref{gaussian}.}
\label{fred}
 \end{figure}

For a given radio survey, the simulated light curve has implications on the part of parameter space that the survey probes. One of the clearest ways to examine these implications is by looking at probability contour plots. Figures~\ref{tophat}-\ref{fred} show the probability contours for example light curves included in the simulations code. The horizontal axis shows the characteristic duration of the transient, which is defined slightly differently for each light curve shape: the tophat and parabolic light curves' characteristic duration is the duration that the transient's flux is non-zero; the characteristic duration for the Gaussian is the standard deviation; and the duration of the FRED, Wilma, and ERED light curves is the e-folding time. The vertical axis shows the characteristic flux, which is the peak flux for all light curves that are currently implemented (but could vary for more exotic light curve shapes). The color legend shows the probability of detecting a source as a transient at a given duration and flux. Note that a source that is detected in every observation would not be a transient. A probability of 1 means that the survey detects every transient source at the particular flux and duration. Note how the region where the transient is always detected changes for the different light curve shapes. The reason for some of these differences is discussed in detail in section~\ref{lcdiscussion}.

   

\subsection{Main Detection Algorithm}

In the transient simulations, a large number of sources need to be generated based on the user's settings. Parameters such as the source flux, duration, and the start, end or critical time (depending on the light curve type) are generated in a uniformly random fashion in log10 space via the random number generator in Numpy. 

The main detection algorithm tests whether or not the simulated sources will be detected in the observations. For this step, the code iterates over each observation, calculating the integrated flux for all the simulated sources, and testing if these integrated fluxes are greater than the sensitivity of the observation multiplied by a user-specified detection threshold. After this detection step, the sources that are detected in every observation are removed from the detection list, since they are constant sources and not transients. 

The number of detected transient sources together with the number of simulated sources are used to generate probabilities of detection for each flux and duration bin. Assuming that transients are distributed as a Poisson distribution, the probabilities are used to calculate limits on transient surface densities and rates. In the case of no transient detections in a survey, the Poisson probability mass function can be inverted to give an upper limit. In case of transient detections in the survey, the code uses the $\chi^2$ distribution \citep{12005udd3.inbook.....JKK} to calculate the upper and lower limits on the transients rates, by inputting the user-provided confidence level and the number of transient detections in the survey. 



\subsection{Gaps}

An important ingredient in calculating transient rates accurately is taking into account gaps of varying sizes during observations and surveys. These gaps may exist for a variety of reasons. In the case of radio observations, a long observation on a particular source has to be broken up into scans that are briefly interrupted by observations of a calibrator source. For measuring the flux of a particular source of interest, these gaps are usually unimportant, but for the purposes of calculating transient rates, especially transient rates in a regime where the transients may be shorter than the size of the gaps, it is important to account for these gaps. 

Gaps are accounted for in the simulations code base by specifying a gaps file. This file contains all the sub-observations, also known as scans, that make up the full length observation. By running the simulations over the scans, and averaging together the measured flux in each scan, we are able to account for realistic gaps in observations. By addressing the issue of gaps in observations, this allows the transient rate calculations to account for multiple different timescales and different sensitivities present in the same survey in an accurate way that would not be possible, or at least very challenging, to do in an analytic fashion \citep{2016MNRAS.459.3161C}.

\subsection{False Detections}
False transient detections is an issue that affects real transient searches and should therefore be included in transient simulations. When an astronomical source is close to the detection threshold, any small amount of measurement error, either statistical or systematic, can change it from a detection to a non-detection or vise-versa. Since this can be true for every observation in the survey, there can exist a fairly wide distribution of false transient detections, governed by the sensitivities of the observations. These sources will be flagged as transients, which is an issue because they are not real transients but merely faint sources of constant flux. Therefore, it is necessary to find a way to eliminate these sources from consideration as transients. 

In order to solve this problem, a second run through the detection algorithm is performed using sources with tophat light curves along with duration and start times that ensure that they ought to be constant sources. After the false detections of these constant sources are calculated, the number of sources detected are counted from the minimum flux simulated until 99\% of the falsely detected sources are accounted for. At the flux level where 99\% is reached, we define this to be the false detection limit. This is shown in all of the probability contour plots, such as in figures~\ref{tophat}-\ref{fred}, and transient rate plots as a horizontal line.

\subsection{Multiple Pointings}
Real surveys can involve multiple pointings that overlap, resulting in an uneven probing of the sky. This creates opportunities and challenges for determining transient rates in these regions of the sky, due to the differences in timescales and observed area. 
In order to account for this, the simulations accurately calculate the area of each region on the sky, and then determine the transient rates for each region on the sky. This is currently implemented for a maximum of three overlapping pointings with possibly varying observing timescales and cadences, but an expansion of this is easily doable. 

\begin{figure}[b]
\includegraphics[scale=0.5]{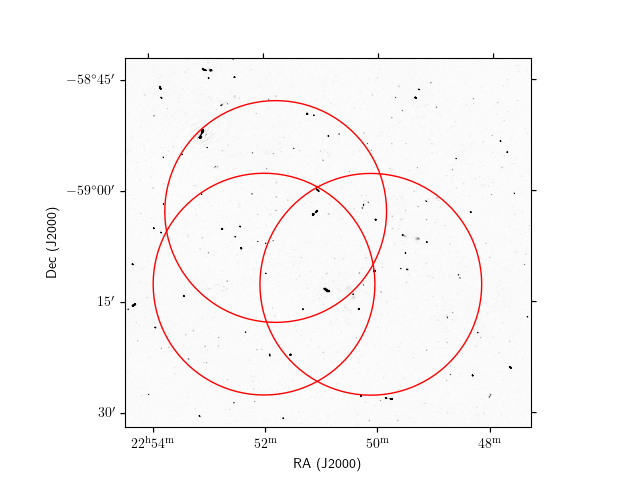}
\caption{An example of three overlapping pointings with red circles representing three different telescope pointings that overlap}
\label{threepointings}
 \end{figure}

Figure \ref{threepointings} shows a simulated example of three overlapping pointings. Each red circle represents a pointing of the telescope. It can be seen that there are also three double overlap regions and one triple overlap region. For each of these regions, the transient rates will be different due to the differences in observing cadence and time.

\section{Results and Discussion}\label{sec:results}

\subsection{A Realistic Survey}

\begin{figure*}[h]
\includegraphics[scale=0.35]{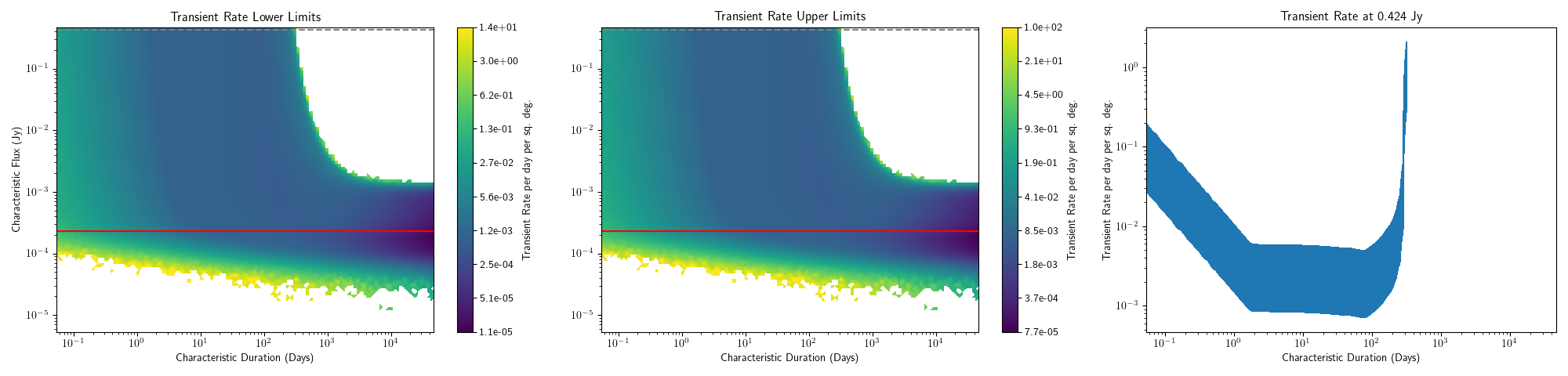}
\caption{Lower limits (left) and upper limits (middle) on transient rate for a realistic survey setup. The red line indicates the false detection limit. The plot on the right shows the limits on the transient rate at 0.424 Jy, which is marked on the two plots to the left with a dashed line at the very top of the plots.}
\label{realscen}
 \end{figure*}

For the purpose of demonstrating the capabilities of the transient simulations code, a survey setup similar to that in \citet{10.1093/mnras/stz3027} is used: 46 weekly observations of 13 minutes in duration, with the rms noise of the observations being varied as a Gaussian with mean 35 $\mu$Jy and standard deviation 5 $\mu$Jy and a detection threshold of $5\sigma$. Given that transients were detected in \citet{10.1093/mnras/stz3027} , we also demonstrate the ability of the simulations code to calculate transient rates based on detections. Assuming Poisson statistics, one can calculate the upper and lower limits on the transient rate, as explained in the previous section. In the configuration for this simulations run, two transient detections are used as input to calculate the rates along with a 95\% confidence interval. The light curve type used for this example is the Gaussian.

Figure \ref{realscen} shows the results of these simulations. The left-side plot shows the lower limits on the transient rate, the middle plot shows the upper limits on the transient rate, and the plot on the right shows the example of transient rate limits at 0.424 Jy as a function of transient duration. The horizontal red lines indicate the 99\% false detection rate. These plots show the transient rate limits: we can see that at 0.424 Jy and around a transient duration of 10 days, the transient rate is between $7\times10^{-4}$ and $5\times10^{-3}$ transients per day per square degree. 

\subsection{Light Curves}\label{lcdiscussion}
Figures~\ref{tophat}-\ref{fred} shows the probability contours for the tophat, parabolic, Gaussian, and FRED light curves. These probabilities are the ratios of the detections to total simulated sources for each duration and flux bin. Each plot has a region of parameter space where all of the transients are detected. As shown by \citet{2017MNRAS.465.4106C}, in the tophat case this is bounded on the left by the duration of the longest gap between consecutive observations. The boundary on the right corresponds to the longest possible duration transient that will still be considered a transient and not a constant source. In other words, this duration is slightly less than the length of the entire survey, since a transient of this length would be detected in every observation except for one.  For the FRED, Gaussian, Wilma, and ERED light curves, we observe that the boundaries around this same region are curves. In~\ref{sec:lc:appendix}, we go into detail on finding the equations for these curves. 

Examining the probability contours for the parabolic light curve in figure~\ref{parabolic} shows a plot that looks closer to the tophat than the other light curves due to the vertical boundaries on the region where the probability is equal to 1. While this may seem counter-intuitive, a similarity between the parabolic and tophat light curve is that they both have a fixed start and stop time at which the flux drops to zero. All the other example light curves approach but never reach zero. For this reason, we use a value to characterize the duration such as the e-folding time for the FRED, or the standard deviation in the case of the Gaussian light curve. Using these values to characterize the duration is what causes the difference in these probability contours. As an example, for the FRED light curve, if the transient has a low flux compared to the sensitivity of the observations, then the duration of the transient that would be detected might be something closer to its e-folding time. In contrast, a very bright transient would be detected well past its e-folding time. This is the reason why these light curves seem to curve away to the left as flux increases in these probability contour plots: the actual duration that is detected in the survey becomes longer. If we were able to define the duration of the transient by the duration that is actually detectable in the survey, then we could make all of the probability contour plots have the kinds of vertical boundaries that we see in the tophat and parabolic light curves. However, defining the durations this way, would make the simulations much more computationally and mathematically complex to the extent that it makes this prohibitive. 

\subsection{False Detections}  \label{fddiscussion}
\begin{figure}[h]
\includegraphics[scale=0.5]{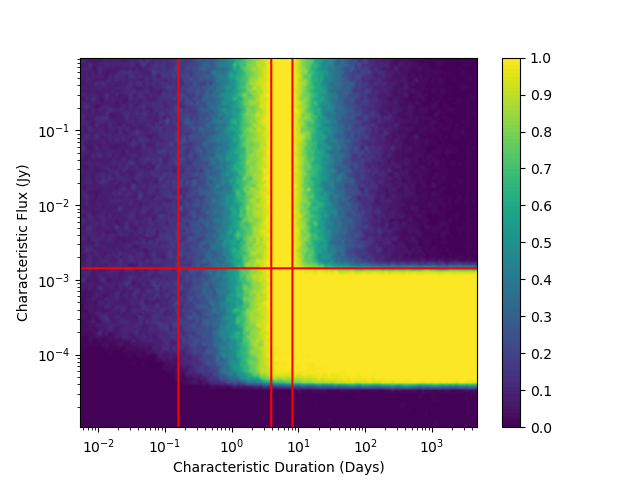}
\caption{Probability contours for a tophat light curve in a survey with very different observation sensitivities (see main text for details).}
\label{fig9}
 \end{figure}

\begin{figure}[h]
\includegraphics[scale=0.5]{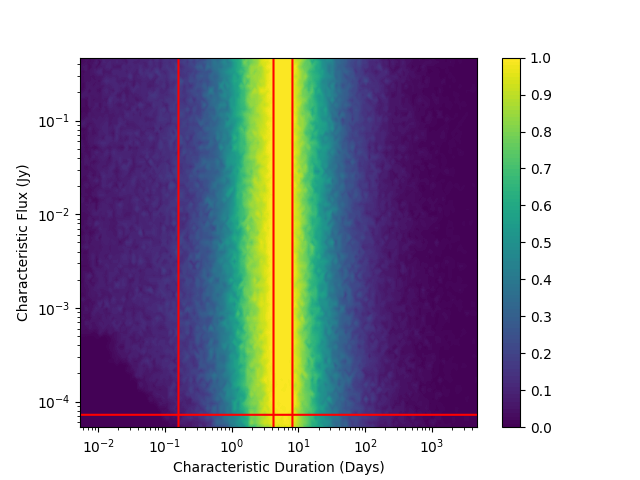}
\caption{Probability contours for a tophat light curve in a survey with only 4 hour observations at an rms noise of around $9~\mu$Jy.}
\label{sample4hr}
 \end{figure}

\begin{figure}[h]
\includegraphics[scale=0.5]{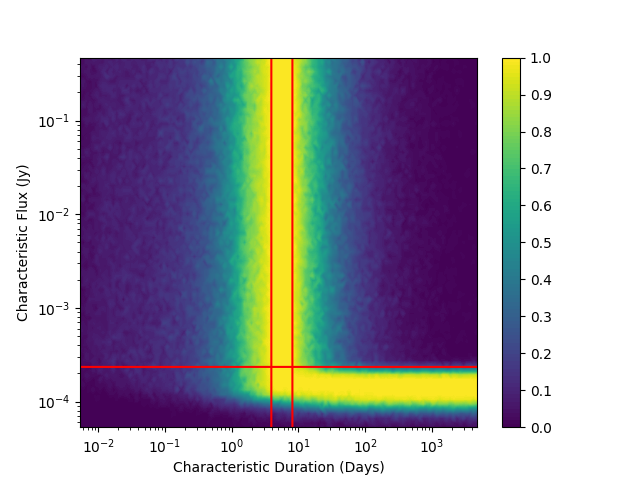}
\caption{Probability contours for a tophat light curve in a survey with only 15 minute observations at an rms noise of around $30~\mu$Jy.}
\label{samplescan}
 \end{figure}

\begin{figure}[h]
\includegraphics[scale=0.5]{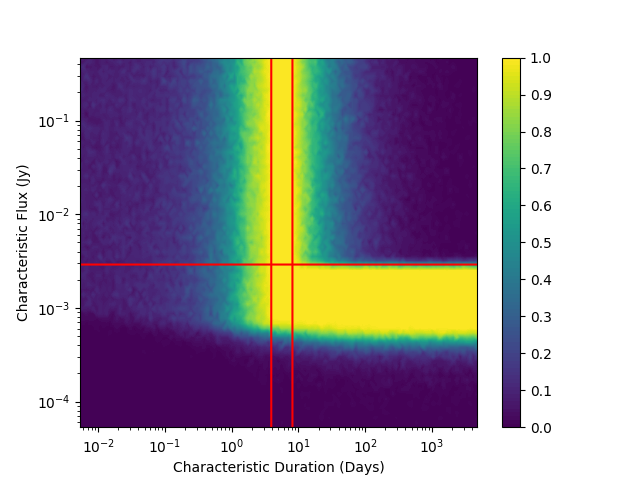}
\caption{Probability contours for a tophat light curve in a survey with only 8 second observations at an rms noise of around $350~\mu$Jy.}
\label{sampleint}
 \end{figure}

Figure~\ref{fig9} shows an example of a survey with a large number of false detections of transients. This can happen when including images that are grouped around very different sensitivity scales. In the example shown here, the survey included observations on three very different time scales and sensitivities: 4 hour observations with an rms noise of around $9~\mu$Jy; 15 minute observations with an rms noise of around $30~\mu$Jy; and 8 second observations with a noise around $350~\mu$Jy. Including all of these images in one run of the simulations creates many false detections. In this example, it is better to run simulations of these three different time and sensitivity scales separately. In figures~\ref{sample4hr}-\ref{sampleint}, the probability contours for the three different timescales are shown separately. From these plots, we can clearly see that the false detection limit is much lower on two of the three timescales and a little higher on the shortest timescale.

\subsection{Gaps} \label{Gaps}
In order to demonstrate the capability of including observations with gaps, an observation file was created with weekly 4 hour observations (instead of 13 minutes), containing gaps within the weekly observation, and a total survey duration of 46 weeks (as in the previous example). The gaps were typical for a target-gain calibration loop in radio observations: 5 minutes on a calibrator field followed by 15 minutes on a science target field. The noise in the 15-minute scans was simulated like before, with a mean of 35 $\mu$Jy and a standard deviation of 5 $\mu$Jy in the target observations. For the full four hour observations, the noise was scaled as $1/\sqrt{time}$ and simulated as a Gaussian with a mean of 8 $\mu$Jy and a standard deviation of 1 $\mu$Jy. Due to the nature of having a bright calibrator source in a field, the noise for the calibrator observations was higher than would be suggested by scaling by $1/\sqrt{time}$. The noise of the 5 minute scans of the calibrator observation was simulated to be 100 $\mu$Jy with a standard deviation of 15 $\mu$Jy. The noise of the combined image of the calibrator scans was simulated to be 25 $\mu$Jy with a standard deviation of 4 $\mu$Jy. For this example we assume that there are no detected transients in this simulated survey.


\begin{figure}[ht]
\includegraphics[scale=0.5]{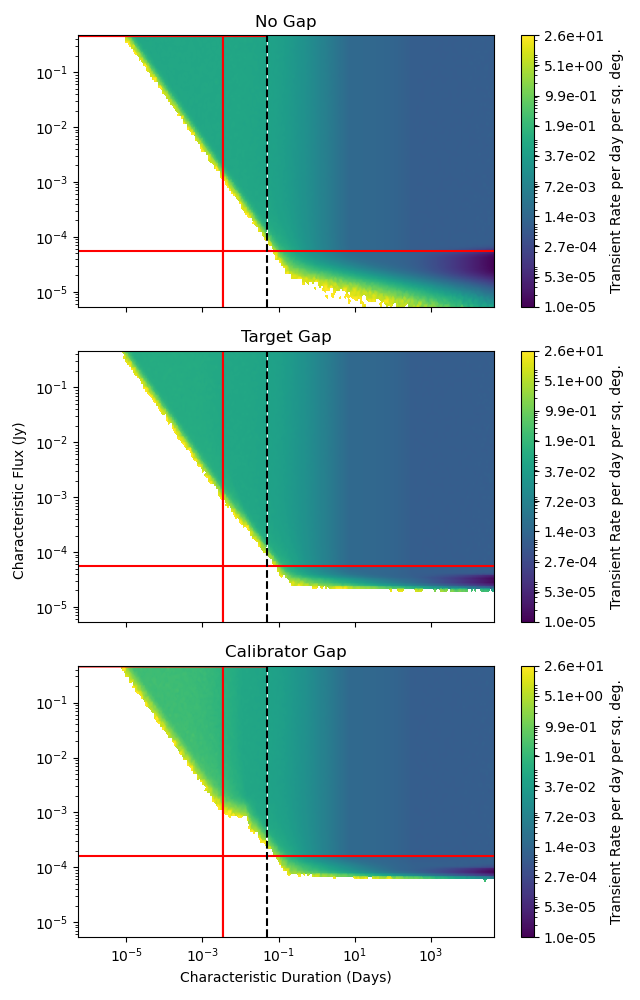}
\caption{Upper limits on transient rate for transients with a tophat light curve in a survey with 4 hour observations with no gaps (top), 5-minute gaps in between 15-minute observations (middle), and 15-minute gaps in between 5-minute observations (bottom); see main text for sensitivities of observations. The dashed black line marks where two different simulations were combined into a single plot. The vertical red line indicates 5 minutes, and the horizontal red line indicates the false detection limit.}
\label{fig3}
 \end{figure}

\begin{figure}[ht]
\includegraphics[scale=0.5]{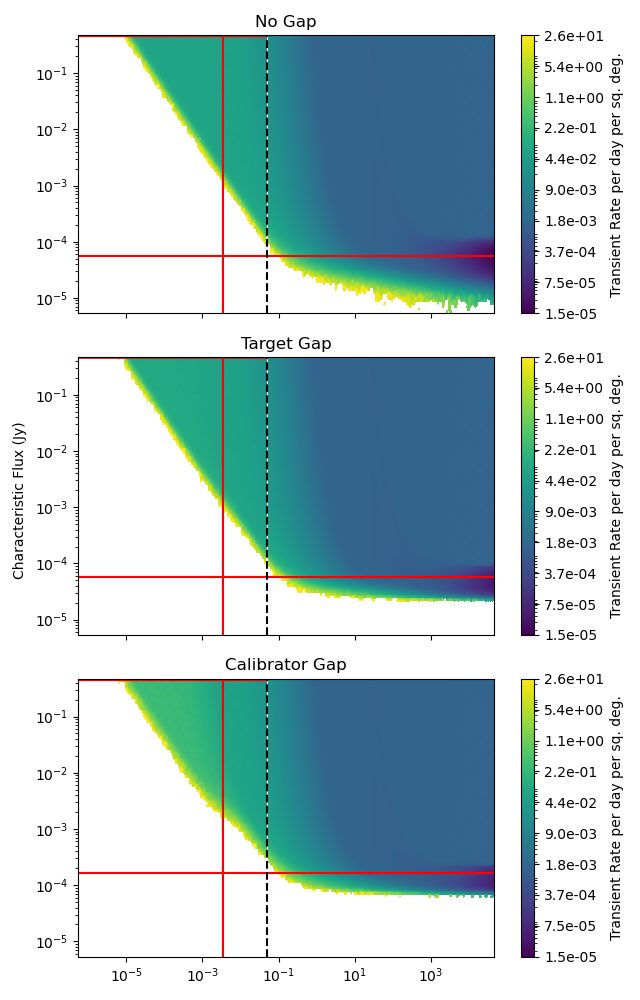}
\caption{Upper limits on transient rate for transients with a FRED light curve in a survey with 4 hour observations with no gaps (top), 5-minute gaps in between 15-minute observations (middle), and 15-minute gaps in between 5-minute observations (bottom); see main text for sensitivities of observations. The dashed black line marks where two different simulations were combined into a single plot. The vertical red line indicates 5 minutes, and the horizontal red line indicates the false detection limit.}
\label{fig4}
 \end{figure}

\begin{figure}[h]
\includegraphics[scale=0.5]{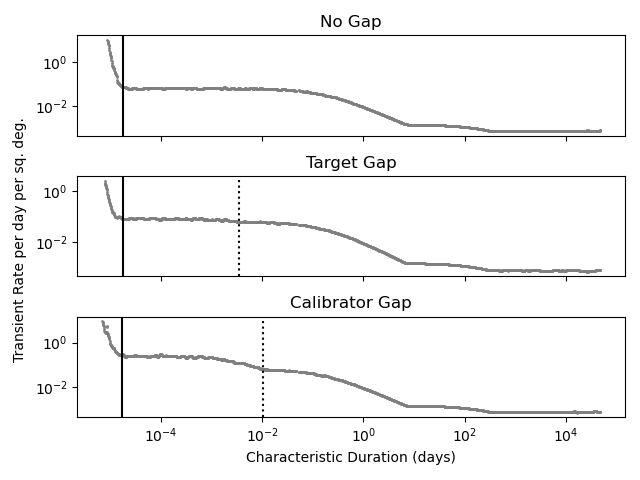}
\caption{Upper limits on transient rates for transients with a tophat light curve in a survey with 4 hour observations with no gaps (top), 5-minute gaps in between 15-minute observations (middle), and 15-minute gaps in between 5-minute observations (bottom), for transients with a peak flux of 0.464 Jy.}
\label{fig5}
 \end{figure}

\begin{figure}[h]
\includegraphics[scale=0.5]{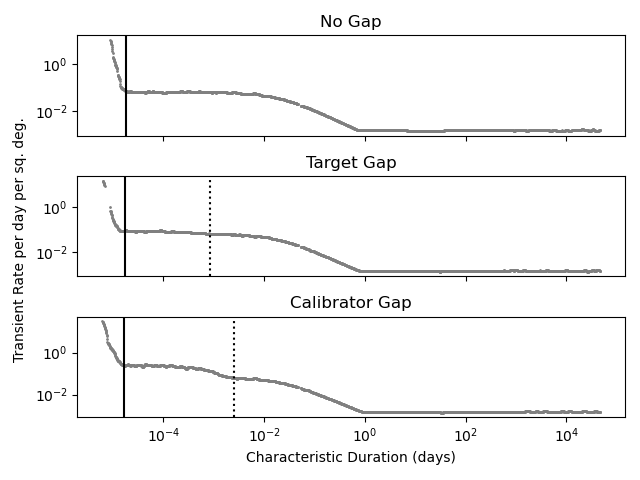}
\caption{Upper limits on transient rates for transients with a FRED light curve in a survey with 4 hour observations with no gaps (top), 5-minute gaps in between 15-minute observations (middle), and 15-minute gaps in between 5-minute observations (bottom), for transients with a peak flux of 0.464 Jy.}
\label{fig6}
 \end{figure}

\begin{figure}[h]
\includegraphics[scale=0.5]{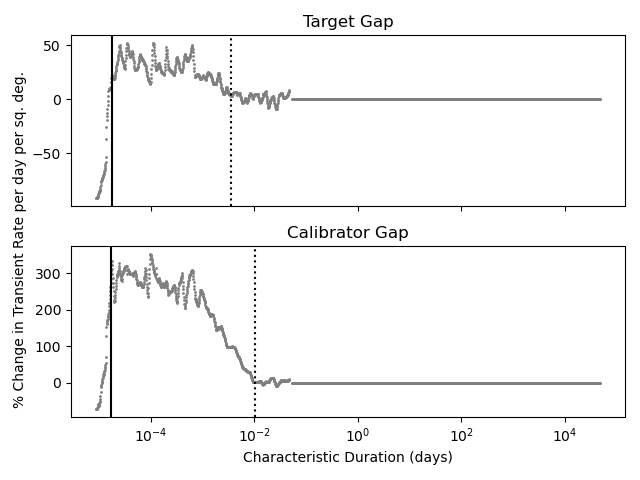}
\caption{Percent difference in upper limits on transient rates for transients with a tophat light curve for transients with a peak flux of 0.464 Jy, from no gaps in a survey with 4 hour observations to 5-minute gaps (top) and 15-minute gaps (bottom).}
\label{fig7}
 \end{figure}

\begin{figure}[h]
\includegraphics[scale=0.5]{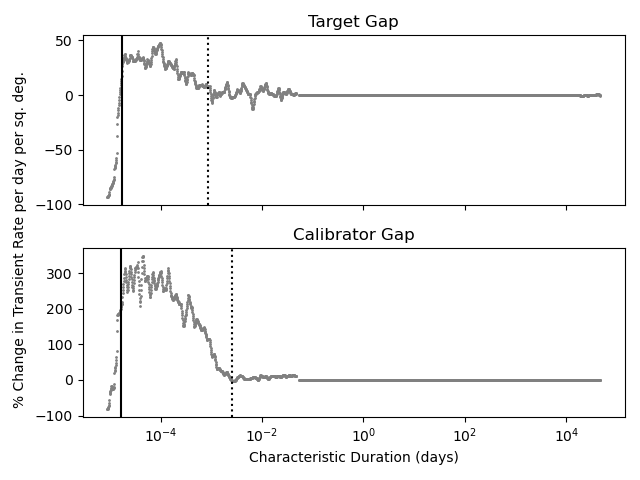}
\caption{Percent difference in upper limits on transient rates for transients with a FRED light curve for transients with a peak flux of 0.464 Jy, from no gaps in a survey with 4 hour observations to 5-minute gaps (top) and 15-minute gaps (bottom).}
\label{fig8}
 \end{figure}

Using these simulations, we can show how accounting for gaps results in more accurate transient rate calculations. This is particularly important on timescales close to the length of the gap itself. In order to test the gaps algorithm, three observational scenarios were used: the calibrator field with 15 minute gaps, the target field with 5 minute gaps, and a full four hour observation with no gaps. These scenarios provide a comparison between different extremes of gaps in observations. These simulations were done with both tophat and FRED light curves. Figures \ref{fig3}-\ref{fig8} show the results of these scenarios. Figures~\ref{fig3} and~\ref{fig4} show upper limits on transient rates in the color legend, with transient duration on the horizontal axis and characteristic flux on the vertical axis. Figure~\ref{fig3} is for a tophat light curve and figure~\ref{fig4} is for a FRED light curve. The three plots in each figure show the difference in transient rates when not accounting for gaps in a weekly survey with 4 hour observations (top), when accounting for 5 minute gaps in a 4 hour science target observation (middle), and when accounting for 15 minute gaps in calibrator observations (bottom). 

In figures~\ref{fig3} and~\ref{fig4} we can see a diagonal trend at the shortest durations below which there are no colored contours. This boundary marks the transients that are shortest in duration and lowest in flux to possibly be detected. It is a diagonal because it is the fluence that determines if a transient is detected \citep{2017MNRAS.465.4106C}; and in the FRED case, for short durations the integrated flux becomes identical to the tophat case. The blank space in the bottom left of the plots represents the region of transient parameter space that cannot be probed by the simulated survey. The red vertical lines on these plots mark 5 minutes, the length of the gaps in the target observations. 

Differences in the transient rate are small and difficult to distinguish between the target gap and no gap plots in figures~\ref{fig3} and~\ref{fig4}. The calibrator gap shows a bit of a departure from the others: examining closely reveals a slightly different trend to the left of the red line for both light curves. This departure from the case of having no gaps or the case of a smaller gap only shows in the part of parameter space that has the smallest duration transients. When transients are longer in duration, they are not likely to fall in the gaps and are more likely to be detected in an observation. 

Figures~\ref{fig5} through~\ref{fig8} show the differences between the different gaps in a different way. Figures~\ref{fig5} and~\ref{fig6} show the transient rates from figures~\ref{fig3} and~\ref{fig4} on the vertical axis at a constant flux of 0.464 Jy. Figures~\ref{fig7} and~\ref{fig8} show the percent difference in transient rate between the gaps and no gaps cases. The top panel of figures~\ref{fig7} and~\ref{fig8} shows the difference between the target gap and no gap cases, and the bottom plot shows the difference between the calibrator gap and no gap cases. As we can see there is an appreciable difference when accounting for 5 minute gaps in a target observation, and a significant difference of nearly 300\% when accounting for 15 minute gaps in the calibrator observation.

 \begin{figure*}[h]
\includegraphics[scale=0.38]{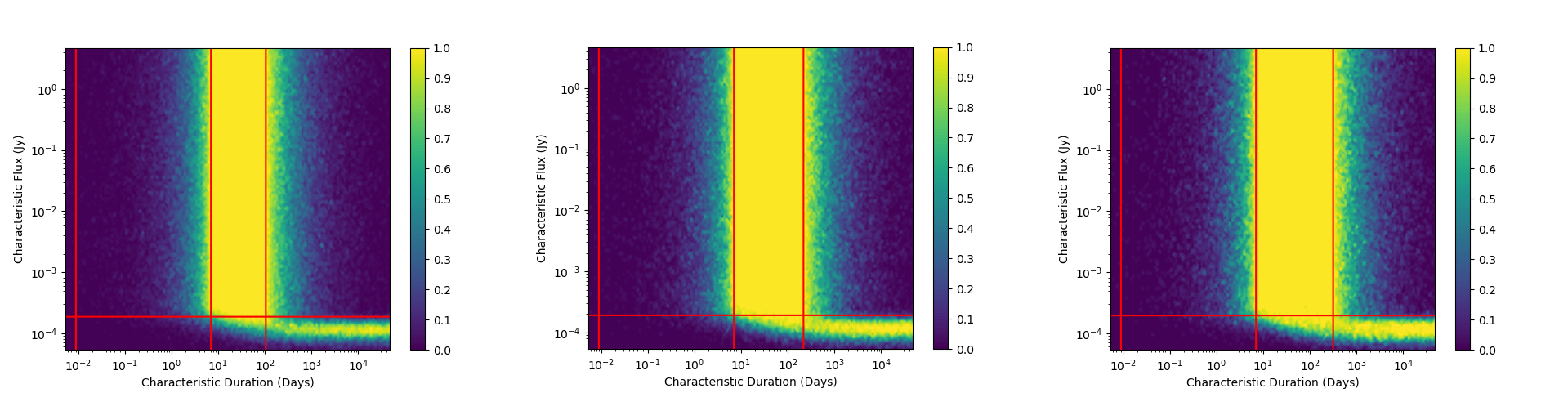}
\caption{A probability contour plot for a tophat light curve in a region with no overlap (left), double overlap (center), and triple overlap (right), as seen in figure~\ref{threepointings}. } 
\label{threeregion}
 \end{figure*}

\subsection{Multiple Pointings}
Calculating transient rates for multiple overlapping pointings gives a more complete picture of how a survey can probe transient parameter space. An example of such a survey is used here and illustrated in figure~\ref{threepointings}: three circular fields of view each with a radius of 1.4 degrees. The details are summarized in tables~\ref{simsurvalt} and~\ref{simsurvseq} below. This setup has seven different regions: three that are probed only by one of the pointings, three that are probed by two pointings, and one that is probed by all three pointings. The three different fields may be observed at various cadences that affect the transient rates in the different areas. If one calculates the probability contours for a tophat transient, as is shown in figure~\ref{threeregion}, one can compare the single pointing (left) with a double overlapping pointing (middle) and a triple overlapping pointing (right). Note how regions with more overlap have a larger region in the transient duration space where the probability of detecting the transient is equal to 1. 

\begin{figure*}
\includegraphics[scale=0.45]{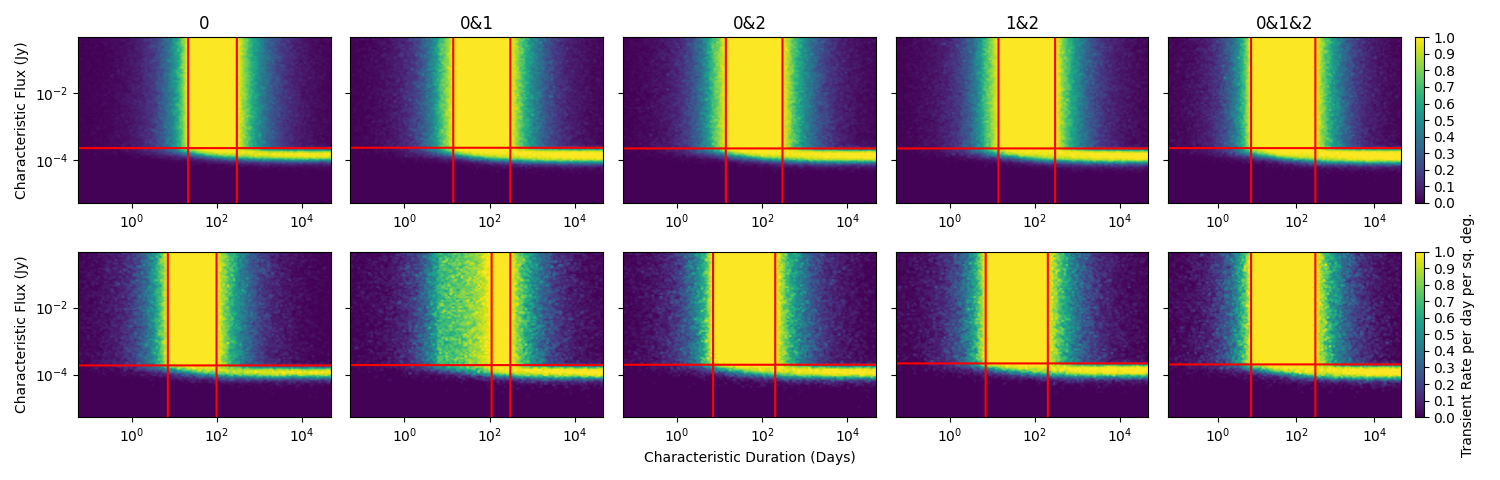}
\caption{Probability contours for all of the regions in survey setup 1 (top) and survey setup 2 (bottom). Each pointing is labeled 0, 1 or 2 and the overlap between two pointings are indicated with an `\&', for example 0\&1.}
\label{multirgnprob}
 \end{figure*}

For a comparison of different survey cadences, one example survey shown in table~\ref{simsurvalt} alternates between each of the three pointings each week, and another one shown in table~\ref{simsurvseq} is set up to observe each pointing exclusively before moving to the next one. Figure~\ref{multirgnprob} shows that the probability contours for the triple overlapping region, labelled 0\&1\&2, are the same for both scenarios, as expected. We do, however, see slight differences in the regions with no overlapping pointings in the part of parameter space where transients are best detected. This difference is due to the variations in the maximum gap and survey length in the two survey setups. One particular region with two overlapping pointings, labelled 0\&1, shows the most striking differences between the survey setups. Survey setup 1 produces two of the double overlap regions with good limits on transient rate and one double overlap region with poor limits on the transient rate. In this case, the region that is observed in both the first observed field and the last observed field will have an extremely large gap. Survey setup 2 produces much more consistent detection regions which may suggest that it is the better choice if more uniform transient rate limits are the goal. 


\begin{center}
\begin{table*}
\begin{tabular}{ |c|c|c|c|c|c|c|} 
 \hline
 RA (J2000) & DEC & ID & Area (deg$^2$) & Duration (days) & Start (MJD) & End \\
 \hline
274.7345 & 7.7974 & 0 & 6.1572 & 294.01 & 58997.54 & 58703.53 \\
 275.0913 & 7.1851 & 1 & 6.1572 & 294.01 & 59011.54 & 58717.53 \\
 275.4482& 7.7974 & 2 & 6.1572 & 294.01 & 59004.54 & 58710.53 \\
 274.9130 & 7.4913 & 0\&1 & 4.1987 & 308.01 & 59011.54 & 58703.53 \\
 275.0913 & 7.7975 & 0\&2 & 4.1987 & 301.01 & 59004.54 & 58703.53 \\
 275.2696 & 7.4913 & 1\&2 & 4.1987 & 301.01 & 59011.54 & 58710.53 \\
 275.0913 & 7.5926 & 0\&1\&2 & 3.4360 & 308.01 & 59011.54 & 58703.53 \\
 \hline
\end{tabular}
\caption{Simulated survey alternating between pointings weekly}
\label{simsurvalt}
\end{table*}

\begin{table*}
\begin{tabular}{ |c|c|c|c|c|c|c|} 
 \hline
 RA (J2000) & DEC & ID & Area (deg$^2$) & Duration (days) & Start (MJD) & End \\
 \hline
274.7345 & 7.7974 & 0 & 6.1572 &  98.16 & 58801.70 & 58703.53 \\
 275.0913  & 7.1851 & 1 & 6.1572 &  98.16 & 59011.70 & 58913.53 \\
 275.4482 & 7.7974 & 2 & 6.1572 &  98.16 & 58906.70 & 58808.53 \\
 274.9130 & 7.4913 & 0\&1 & 4.1987 & 308.16 & 59011.70 & 58703.53 \\
 275.0913  & 7.7975 & 0\&2 & 4.1987 & 203.16 & 58906.70 & 58703.53 \\
 275.2696 & 7.4913 & 1\&2 & 4.1987 & 203.16 & 59011.70 & 58808.53 \\
 275.0913  & 7.5926 & 0\&1\&2 & 3.4360 & 308.16 & 59011.70 & 58703.53 \\
 \hline
\end{tabular}
\caption{Simulated survey moving between pointings sequentially}
\label{simsurvseq}
\end{table*}
\end{center}

\section{Performance}\label{performance}
\begin{figure}[h]
\includegraphics[scale=0.5]{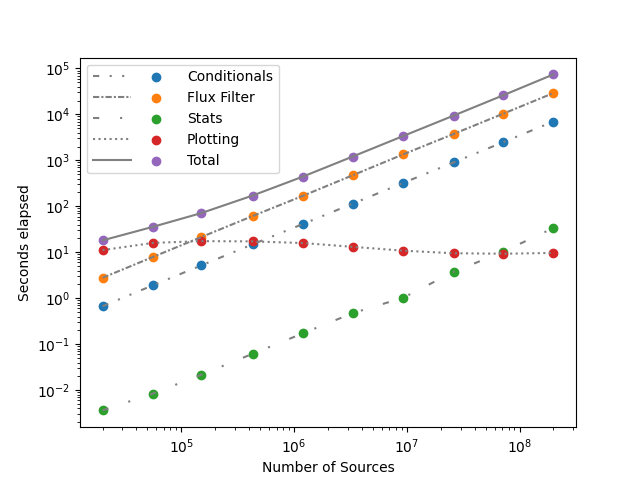}
\caption{Execution time as a function of the number of simulated sources.}
\label{fig10}
 \end{figure}

\begin{figure}[h]
\includegraphics[scale=0.5]{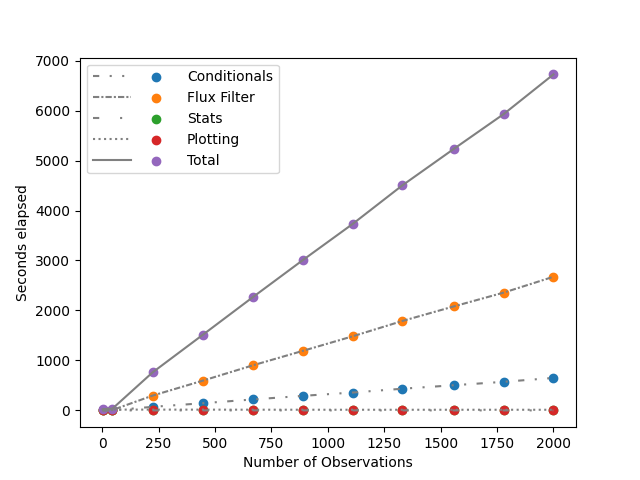}
\caption{Execution time as a function of the number of observations.}
\label{fig11}
 \end{figure}

Figure~\ref{fig10} shows how the simulations scale in execution time as a function of the number of sources simulated, for the example of 46 observations of 13 minutes. Figure~\ref{fig11} shows scaling in execution time as a function of the number of observations when the number of sources is held constant at $4.3\times10^{5}$.  All of the simulations for this example were performed on a 2020 Apple Macbook Pro 13 inch model with the M1 chip. In addition to showing the total execution time, key portions of the code are shown as well. The conditionals and flux filtering steps take place when the algorithm is determining if the sources are detected in observations. These steps are so-named because they filter sources based on the calculated integrated flux, and do a large number of boolean and bit operations to calculate and store each transient's state as either detected or not-detected. The stats step is the step that aggregates the detections into probabilities. Finally, the plotting step is where all of the detection statistics, observation information, and false detection information is plotted. Since the data is broken down into a grid in order to plot, the plotting step has no scaling with the number of sources, so it takes a constant amount of time. 

Not shown here are the impacts of a few features and algorithms. The false detection algorithm re-runs the conditionals, flux filtering, and stats steps. In the case of a tophat light curve, this means the false detection algorithm would slightly less than double the amount of time (the plotting step is not doubled). In the case of other light curves, it could have a different impact, usually lower, since the false detection algorithm always uses tophat light curves to simulate constant sources. Another factor not shown here is the impact of having multiple pointings. For example, in a survey setup with two overlapping pointings there will be three regions. Therefore, the run time will be about three times longer than for a single region, assuming equal numbers of observations in each region. 

%

\section{Future Applications} \label{futureapplications}
\citet{2017MNRAS.465.4106C} shows the existence of a region of 100\% detection probability for the FRED and tophat light curves. The work presented here shows that this region of 100\% detection can be found in a wide variety of light curves. Optimizing the survey parameters to maximize this region of 100\% detection has many potential applications for future surveys. In addition, transient searches can use the false detection calculations to decide the best manner to plan a transient search in a given survey. The wide variety of data outputs can assist with a number of scientific goals that one might have for a survey, and the flexibility of the code can be easily adapted to those purposes.

One application of the code is using the transient simulations to optimize resource allocation. Such optimizations can be done both in terms of survey design and also the data reduction after the observation itself. Simulating a variety of survey setups, such as in section~\ref{Gaps}, is one way to ensure that the survey will accomplish it goals most effectively. In addition, simulations can also be done for varying aspects related to the data reduction, such as the timescale that images are made on. By simulating a combining or splitting up of the observations in a survey into multiple different timescales, one can find an optimal way to search for transients on these multiple time scales with a minimum of re-imaging. These optimizations are becoming increasingly more important with radio facilities such as MeerKAT \cite{2016mks..confE...1J} and the LOw-Frequency Array (LOFAR) \cite{2013A&A...556A...2V} which can easily use terabytes of disk space and considerable other computer resources as well. The Square Kilometer Array \cite{2009IEEEP..97.1482D} and next-generation Very Large Array \cite{2018ASPC..517....3M}, and other upcoming facilities, will surely require even more resources.

The previously mentioned tools for planning surveys have potential to be expanded to be even more helpful in the future. A potential future update to these simulations could include a tool to help calculate optimal pointings for smoothly probing a large area of the sky in both space and time. 

The simulations code presented in this paper accounts for a large number of realistic effects that complicate transient searches and calculate transient rates from surveys. For research into particular kinds of sources, future upgrades can be made for particular light curves and population numbers that reflect certain sources of interest. 

Finally, even though this simulations code has been designed for surveys in the radio regime, and the examples in this paper are based on this particular use case, it can easily be adapted and applied to other spectral regimes.

\section{Conclusions} \label{conclusions}
Simulating transients, following the methodology and code case presented here, allows for calculating transient rates that are highly accurate due to the implementation of a variety of observational effects. The simulations presented here account for a variety of observing sensitivities, pointings, survey cadences, and gaps within observations and surveys. Furthermore, it has been made easy to obtain, since it will be freely available for download through Github, and easy to use through the use of a modular design, the inclusion of scripts to extract metadata from observations, and updates for modern versions of Python.

\section{Acknowledgements} 

The authors would like to thank the referee for their constructive comments that helped improve the paper.
The authors would like to acknowledge the ThunderKAT collaboration for the valuable sharing of knowledge and resources, and Michael Moss for his helpful comments and feedback on this manuscript. 
This work was completed in part with resources provided by the High Performance Computing Cluster
at The George Washington University, Information Technology, Research Technology Services.





\appendix
\section{Included Light Curves}
\label{sec:lc:appendix}

Here we present and briefly discuss the light curve shapes that are currently included in the simulations code base.

\subsection{Tophat}

The tophat is the simplest transient light curve in concept:
\[F=F_{pk}~\text{for}~t_{start} \le t \le t_{end}\]
It is simply at the peak flux for the entire duration of the transient. The probability contour plot shown in figure~\ref{tophat} has a region in parameter space in which all transients are always detected, which can be referred to as a region of guaranteed detection. This region has vertical boundaries that can be found to have a quite straightforward interpretation \citep{2017MNRAS.465.4106C}. The left-most boundary is the longest gap in the observations or, in other words, the longest duration of a tophat transient that could go undetected. The right-most vertical bounding line corresponds to the longest time scale that a transient could have while not being detected as a constant source. This quantity is the duration of the entire survey minus either the first or last observation. 

\subsection{Fast Rise Exponential Decay}
The fast rise exponential decay (FRED) light curve is defined as instantaneously rising to the peak flux and exponentially decaying with an characteristic duration $\tau$, defined as its e-folding time:
\[F=F_{pk}\,\exp\left[\frac{-(t-t_{start})}{\tau}\right]~\text{for}~t\ge t_{start}\]

This light curve produces a slightly different probability contour, seen in figure~\ref{fred}, in which the bounding lines for the region of guaranteed detection can be interpreted as follows. The left boundary corresponds to the boundary due to the longest gap, like the tophat. However, unlike the tophat, the flux of the FRED light curve approaches but never actually reaches zero as time progresses. Therefore, brighter transients can be detected for longer than the characteristic duration of the transient, making this boundary a curve instead of a vertical line. The boundary condition can be expressed as $F_{int}=S_{gap}$, i.e. the integrated flux of the transient needs to be equal to the sensitivity of the observation the transient would be detected in, which would be the observation after the gap. We can find the integrated flux:
\[F_{int} = F_{pk}\,\tau\,\frac{\exp\left[-\frac{\max(T_{start},t_{start})}{\tau}\right] - \exp\left[\frac{T_{end}}{\tau}\right]}{T_{end}-T_{start}}\]
Since we consider the case where the transient starts in the gap, the start of the observation that detects the transient is equal to the length of time from the start of the transient until the end of the gap, which we label $T_{gap}$: $T_{start}=T_{gap}$. 
Therefore, $T_{end}=T_{start}+\Delta T_{gap}$, where $\Delta T_{gap}$ is the duration of the observation. Inserting the integrated flux into the previous equation and solving for $F_{pk}$ yields:
\[F_{pk}(\tau) = \frac{S_{gap}\,\Delta T_{gap}}{\tau\,\left(\exp\left[-\frac{T_{gap}}{\tau}\right] - \exp\left[\frac{(T_{gap} + \Delta T_{gap})}{\tau}\right]\right)}\]

The right boundary is the boundary for the longest timescale. We can follow the same procedure as the left boundary, finding that $S_{obs} = S_{last}$, the sensitivity of the last observation in the survey. We also find the following modifications:
\[T_{start} = \tau_{survey}  - \Delta T_{last}\]
\[T_{end} = \tau_{survey}\]
\[F_{pk}(\tau) = \frac{S_{last}\,\Delta T_{last}}{\tau\,\left(\exp\left[-\frac{(\tau_{survey} - \Delta T_{last})}{\tau}\right] - \exp\left[\frac{\tau_{survey}}{\tau}\right]\right)}\]

\begin{figure}[h]
\includegraphics[scale=0.5]{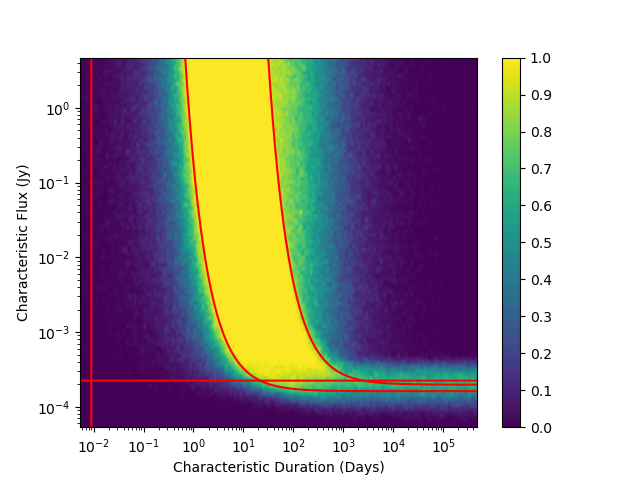}
\caption{Probability contours for the Wilma light curve}
\label{wilma}
 \end{figure}
 
\subsection{Exponential Rise Fast Decay (Wilma)}
In light of the FRED light curve, a natural extension would be to examine the reverse FRED light curve. The light curve ends at the peak flux and has no definite start:
\[F=F_{pk}\,\exp\left[{\frac{(t-t_{end})}{\tau}}\right]\text{ for }t\le t_{end}\]
The probability contour plot for this light curve is shown in figure~\ref{wilma}. As one can see, it is identical to the FRED light curve in figure~\ref{fred}. This makes sense when one realizes that if the entire survey were time-reversed, the light curve would be a FRED. For this reason, the lines bounding the region of guaranteed detection are the same as for the FRED light curve. 

\begin{figure}[h]
\includegraphics[scale=0.5]{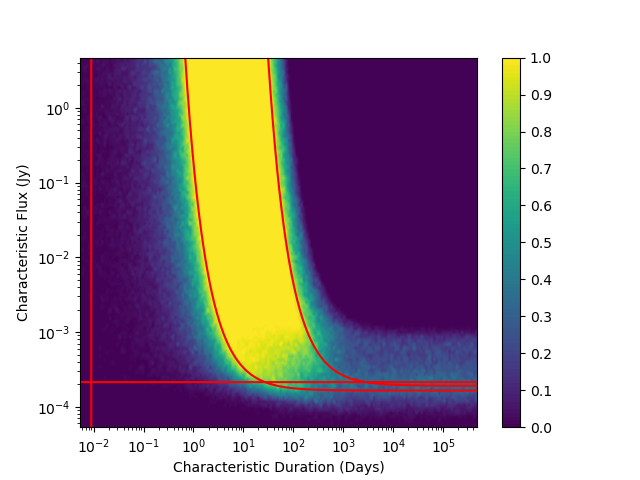}
\caption{Probability contours for the ERED light curve}
\label{ered}
 \end{figure}
 
\subsection{Exponential Rise Exponential Decay}
The Exponential Rise Exponential Decay (ERED) light curve (figure~\ref{ered}) does not have a definite beginning nor end, only a characteristic time at which the flux is the peak flux and $\tau$, which is its e-folding time:
\[F=F_{pk}\,\exp\left[{\frac{(t-t_{char})}{\tau}}\right]\text{ for }t< t_{char}\]
\[F=F_{pk}\text{ for }t=t_{char}\]
\[F=F_{pk}\,\exp\left[{\frac{-(t-t_{char})}{\tau}}\right]\text{ for }t> t_{char}\]
The transients from this light curve also behave similarly to the previous two cases, since this light curve is a Wilma light curve immediately followed by a FRED light curve.  Therefore, if we use the integrated flux to find the curve marking the boundary corresponding to the shortest duration transient that will always be detected, we find:
\[F_{pk}(\tau) = \frac{ S _{gap}\,\Delta T_{gap}}{\frac{\tau}{2}\,\left(\exp\left[-\frac{T_{gap}}{\tau}\right] - \exp\left[\frac{2\Delta T_{gap} + T_{gap}}{\tau}\right]\right)}\]
Similarly, for the flux limit on the longest timescale, we have:
\[F_{pk}(\tau) = \frac{S_{first/last}\,\Delta T_{first/last}}{\frac{\tau}{2}\,\left(\exp\left[-\frac{\tau_{survey}}{\tau}\right] - \exp\left[-\frac{2\Delta T_{first/last} + \tau_{survey}}{\tau}\right]\right)}\]
In this equation, $S_{first/last}$ and $\Delta T_{first/last}$ would correspond to either the first or last observation in the survey depending on which is more sensitive. 

\subsection{Parabola}
Also included is a parabolic light curve defined as follows:
\[F = F_{pk}\,\left(1 - \frac{4}{\tau^2}\left(t - \frac{\tau}{2} - t_{crit}\right)^2\right)\]
$t_{crit}$ is the peak of the light curve, which occurs at half of the duration of the light curve. Since the parabolic light curve starts and ends at zero flux, rather than approaching zero like the exponential or Gaussian light curves, this light curve has a definite duration like the tophat. The light curves with a definite duration all have boundaries in the probability contour plots that are derived in the same way as those for the tophat light curves. The boundaries in the case of the the exponential or Gaussian light curves come about from there being a difference between the characteristic duration and the duration that the transient is actually detected in the observations. 
\subsection{Gaussian}
The final light curve included is a Gaussian-shaped one:
\[F=F_{pk}\,\exp\left[\frac{-(t-t_{crit})^2}{2\left(\frac{\tau}{2}\right)^2}\right]\]
In order to find the boundaries for the region of parameter space where the transients are always detected, we follow the same process as we did for the FRED light curve, and find an equation for the integrated flux (with erf being the Gauss error function): 
\begin{equation*}
\begin{split}
    F_{int} = F_{pk}\,\tau\,\sqrt{\frac{\pi}{8}}\,(\text{erf}\left[\frac{\sqrt{2}(T_{end}-t_{crit})}{\tau}\right]\\
    -\text{erf}\left[\frac{\sqrt{2}(T_{start}-t_{crit})}{\tau}\right])/(T_{end}-T_{start})
\end{split}
\end{equation*}
We can define the boundary on the transient flux needed to be detected as $F_{int}=S_{gap}$.
Using the equation for integrated flux, we find the boundary for the shortest possible duration transient that will always be detected:

\begin{equation*}
\begin{split}
    F_{pk} = \frac{S_{obs}\,\Delta T_{obs}}{\tau\, \sqrt{\frac{\pi}{8}}}\frac{1}{\text{erf}\left[\frac{-\Delta T_{gap}}{\sqrt{2}\,\tau}\right] - \text{erf}\left[\frac{(-2\Delta T_{obs} - \Delta T_{gap})}{\sqrt{2}\,\tau}\right]}
\end{split}
\end{equation*}

We also find the boundary on the right, for the longest possible duration transient before it is considered a constant source:

\begin{equation*}
\begin{split}
    F_{pk} = \frac{S_{obs}\,\Delta T_{obs}}{\tau\, \sqrt{\frac{\pi}{8}}}\frac{1}{\text{erf}\left[\frac{-(\Delta T_{survey} + 2\Delta T_{obs})}{\sqrt{2}\,\tau}\right] - \text{erf}\left[\frac{-\sqrt{2}\,\Delta T_{survey}}{\tau}\right]}
\end{split}
\end{equation*}

 \bibliographystyle{elsarticle-num-names} 
 \bibliography{thesis}

\begin{thebibliography}{25}
\expandafter\ifx\csname natexlab\endcsname\relax\def\natexlab#1{#1}\fi
\providecommand{\url}[1]{\texttt{#1}}
\providecommand{\href}[2]{#2}
\providecommand{\path}[1]{#1}
\providecommand{\DOIprefix}{doi:}
\providecommand{\ArXivprefix}{arXiv:}
\providecommand{\URLprefix}{URL: }
\providecommand{\Pubmedprefix}{pmid:}
\providecommand{\doi}[1]{\href{http://dx.doi.org/#1}{\path{#1}}}
\providecommand{\Pubmed}[1]{\href{pmid:#1}{\path{#1}}}
\providecommand{\bibinfo}[2]{#2}
\ifx\xfnm\relax \def\xfnm[#1]{\unskip,\space#1}\fi
\bibitem[{{Ivezi{\'c}} et~al.(2019){Ivezi{\'c}}, {Kahn}, {Tyson}, {Abel},
  {Acosta}, {Allsman}, {Alonso}, {AlSayyad}, {Anderson}, {Andrew}, {Angel},
  {Angeli}, {Ansari}, {Antilogus}, {Araujo}, {Armstrong}, {Arndt}, {Astier},
  {Aubourg}, {Auza}, {Axelrod}, {Bard}, {Barr}, {Barrau}, {Bartlett}, {Bauer},
  {Bauman}, {Baumont}, {Bechtol}, {Bechtol}, {Becker}, {Becla}, {Beldica},
  {Bellavia}, {Bianco}, {Biswas}, {Blanc}, {Blazek}, {Blandford}, {Bloom},
  {Bogart}, {Bond}, {Booth}, {Borgland}, {Borne}, {Bosch}, {Boutigny},
  {Brackett}, {Bradshaw}, {Brandt}, {Brown}, {Bullock}, {Burchat}, {Burke},
  {Cagnoli}, {Calabrese}, {Callahan}, {Callen}, {Carlin}, {Carlson},
  {Chandrasekharan}, {Charles-Emerson}, {Chesley}, {Cheu}, {Chiang}, {Chiang},
  {Chirino}, {Chow}, {Ciardi}, {Claver}, {Cohen-Tanugi}, {Cockrum}, {Coles},
  {Connolly}, {Cook}, {Cooray}, {Covey}, {Cribbs}, {Cui}, {Cutri}, {Daly},
  {Daniel}, {Daruich}, {Daubard}, {Daues}, {Dawson}, {Delgado}, {Dellapenna},
  {de Peyster}, {de Val-Borro}, {Digel}, {Doherty}, {Dubois},
  {Dubois-Felsmann}, {Durech}, {Economou}, {Eifler}, {Eracleous}, {Emmons},
  {Fausti Neto}, {Ferguson}, {Figueroa}, {Fisher-Levine}, {Focke}, {Foss},
  {Frank}, {Freemon}, {Gangler}, {Gawiser}, {Geary}, {Gee}, {Geha}, {Gessner},
  {Gibson}, {Gilmore}, {Glanzman}, {Glick}, {Goldina}, {Goldstein}, {Goodenow},
  {Graham}, {Gressler}, {Gris}, {Guy}, {Guyonnet}, {Haller}, {Harris},
  {Hascall}, {Haupt}, {Hernandez}, {Herrmann}, {Hileman}, {Hoblitt}, {Hodgson},
  {Hogan}, {Howard}, {Huang}, {Huffer}, {Ingraham}, {Innes}, {Jacoby}, {Jain},
  {Jammes}, {Jee}, {Jenness}, {Jernigan}, {Jevremovi{\'c}}, {Johns}, {Johnson},
  {Johnson}, {Jones}, {Juramy-Gilles}, {Juri{\'c}}, {Kalirai}, {Kallivayalil},
  {Kalmbach}, {Kantor}, {Karst}, {Kasliwal}, {Kelly}, {Kessler}, {Kinnison},
  {Kirkby}, {Knox}, {Kotov}, {Krabbendam}, {Krughoff}, {Kub{\'a}nek},
  {Kuczewski}, {Kulkarni}, {Ku}, {Kurita}, {Lage}, {Lambert}, {Lange},
  {Langton}, {Le Guillou}, {Levine}, {Liang}, {Lim}, {Lintott}, {Long},
  {Lopez}, {Lotz}, {Lupton}, {Lust}, {MacArthur}, {Mahabal}, {Mandelbaum},
  {Markiewicz}, {Marsh}, {Marshall}, {Marshall}, {May}, {McKercher}, {McQueen},
  {Meyers}, {Migliore}, {Miller}, {Mills}, {Miraval}, {Moeyens}, {Moolekamp},
  {Monet}, {Moniez}, {Monkewitz}, {Montgomery}, {Morrison}, {Mueller},
  {Muller}, {Mu{\~n}oz Arancibia}, {Neill}, {Newbry}, {Nief}, {Nomerotski},
  {Nordby}, {O'Connor}, {Oliver}, {Olivier}, {Olsen}, {O'Mullane}, {Ortiz},
  {Osier}, {Owen}, {Pain}, {Palecek}, {Parejko}, {Parsons}, {Pease},
  {Peterson}, {Peterson}, {Petravick}, {Libby Petrick}, {Petry},
  {Pierfederici}, {Pietrowicz}, {Pike}, {Pinto}, {Plante}, {Plate}, {Plutchak},
  {Price}, {Prouza}, {Radeka}, {Rajagopal}, {Rasmussen}, {Regnault}, {Reil},
  {Reiss}, {Reuter}, {Ridgway}, {Riot}, {Ritz}, {Robinson}, {Roby}, {Roodman},
  {Rosing}, {Roucelle}, {Rumore}, {Russo}, {Saha}, {Sassolas}, {Schalk},
  {Schellart}, {Schindler}, {Schmidt}, {Schneider}, {Schneider}, {Schoening},
  {Schumacher}, {Schwamb}, {Sebag}, {Selvy}, {Sembroski}, {Seppala}, {Serio},
  {Serrano}, {Shaw}, {Shipsey}, {Sick}, {Silvestri}, {Slater}, {Smith},
  {Smith}, {Sobhani}, {Soldahl}, {Storrie-Lombardi}, {Stover}, {Strauss},
  {Street}, {Stubbs}, {Sullivan}, {Sweeney}, {Swinbank}, {Szalay}, {Takacs},
  {Tether}, {Thaler}, {Thayer}, {Thomas}, {Thornton}, {Thukral}, {Tice},
  {Trilling}, {Turri}, {Van Berg}, {Vanden Berk}, {Vetter}, {Virieux},
  {Vucina}, {Wahl}, {Walkowicz}, {Walsh}, {Walter}, {Wang}, {Wang}, {Warner},
  {Wiecha}, {Willman}, {Winters}, {Wittman}, {Wolff}, {Wood-Vasey}, {Wu},
  {Xin}, {Yoachim}, and {Zhan}}]{2019ApJ...873..111I}
\bibinfo{author}{{\v{Z}}.~{Ivezi{\'c}}}, \bibinfo{author}{S.~M. {Kahn}},
  \bibinfo{author}{J.~A. {Tyson}}, \bibinfo{author}{B.~{Abel}},
  \bibinfo{author}{E.~{Acosta}}, \bibinfo{author}{R.~{Allsman}},
  \bibinfo{author}{D.~{Alonso}}, \bibinfo{author}{Y.~{AlSayyad}},
  \bibinfo{author}{S.~F. {Anderson}}, \bibinfo{author}{J.~{Andrew}},
  \bibinfo{author}{J.~R.~P. {Angel}}, \bibinfo{author}{G.~Z. {Angeli}},
  \bibinfo{author}{R.~{Ansari}}, \bibinfo{author}{P.~{Antilogus}},
  \bibinfo{author}{C.~{Araujo}}, \bibinfo{author}{R.~{Armstrong}},
  \bibinfo{author}{K.~T. {Arndt}}, \bibinfo{author}{P.~{Astier}},
  \bibinfo{author}{{\'E}.~{Aubourg}}, \bibinfo{author}{N.~{Auza}},
  \bibinfo{author}{T.~S. {Axelrod}}, \bibinfo{author}{D.~J. {Bard}},
  \bibinfo{author}{J.~D. {Barr}}, \bibinfo{author}{A.~{Barrau}},
  \bibinfo{author}{J.~G. {Bartlett}}, \bibinfo{author}{A.~E. {Bauer}},
  \bibinfo{author}{B.~J. {Bauman}}, \bibinfo{author}{S.~{Baumont}},
  \bibinfo{author}{E.~{Bechtol}}, \bibinfo{author}{K.~{Bechtol}},
  \bibinfo{author}{A.~C. {Becker}}, \bibinfo{author}{J.~{Becla}},
  \bibinfo{author}{C.~{Beldica}}, \bibinfo{author}{S.~{Bellavia}},
  \bibinfo{author}{F.~B. {Bianco}}, \bibinfo{author}{R.~{Biswas}},
  \bibinfo{author}{G.~{Blanc}}, \bibinfo{author}{J.~{Blazek}},
  \bibinfo{author}{R.~D. {Blandford}}, \bibinfo{author}{J.~S. {Bloom}},
  \bibinfo{author}{J.~{Bogart}}, \bibinfo{author}{T.~W. {Bond}},
  \bibinfo{author}{M.~T. {Booth}}, \bibinfo{author}{A.~W. {Borgland}},
  \bibinfo{author}{K.~{Borne}}, \bibinfo{author}{J.~F. {Bosch}},
  \bibinfo{author}{D.~{Boutigny}}, \bibinfo{author}{C.~A. {Brackett}},
  \bibinfo{author}{A.~{Bradshaw}}, \bibinfo{author}{W.~N. {Brandt}},
  \bibinfo{author}{M.~E. {Brown}}, \bibinfo{author}{J.~S. {Bullock}},
  \bibinfo{author}{P.~{Burchat}}, \bibinfo{author}{D.~L. {Burke}},
  \bibinfo{author}{G.~{Cagnoli}}, \bibinfo{author}{D.~{Calabrese}},
  \bibinfo{author}{S.~{Callahan}}, \bibinfo{author}{A.~L. {Callen}},
  \bibinfo{author}{J.~L. {Carlin}}, \bibinfo{author}{E.~L. {Carlson}},
  \bibinfo{author}{S.~{Chandrasekharan}},
  \bibinfo{author}{G.~{Charles-Emerson}}, \bibinfo{author}{S.~{Chesley}},
  \bibinfo{author}{E.~C. {Cheu}}, \bibinfo{author}{H.-F. {Chiang}},
  \bibinfo{author}{J.~{Chiang}}, \bibinfo{author}{C.~{Chirino}},
  \bibinfo{author}{D.~{Chow}}, \bibinfo{author}{D.~R. {Ciardi}},
  \bibinfo{author}{C.~F. {Claver}}, \bibinfo{author}{J.~{Cohen-Tanugi}},
  \bibinfo{author}{J.~J. {Cockrum}}, \bibinfo{author}{R.~{Coles}},
  \bibinfo{author}{A.~J. {Connolly}}, \bibinfo{author}{K.~H. {Cook}},
  \bibinfo{author}{A.~{Cooray}}, \bibinfo{author}{K.~R. {Covey}},
  \bibinfo{author}{C.~{Cribbs}}, \bibinfo{author}{W.~{Cui}},
  \bibinfo{author}{R.~{Cutri}}, \bibinfo{author}{P.~N. {Daly}},
  \bibinfo{author}{S.~F. {Daniel}}, \bibinfo{author}{F.~{Daruich}},
  \bibinfo{author}{G.~{Daubard}}, \bibinfo{author}{G.~{Daues}},
  \bibinfo{author}{W.~{Dawson}}, \bibinfo{author}{F.~{Delgado}},
  \bibinfo{author}{A.~{Dellapenna}}, \bibinfo{author}{R.~{de Peyster}},
  \bibinfo{author}{M.~{de Val-Borro}}, \bibinfo{author}{S.~W. {Digel}},
  \bibinfo{author}{P.~{Doherty}}, \bibinfo{author}{R.~{Dubois}},
  \bibinfo{author}{G.~P. {Dubois-Felsmann}}, \bibinfo{author}{J.~{Durech}},
  \bibinfo{author}{F.~{Economou}}, \bibinfo{author}{T.~{Eifler}},
  \bibinfo{author}{M.~{Eracleous}}, \bibinfo{author}{B.~L. {Emmons}},
  \bibinfo{author}{A.~{Fausti Neto}}, \bibinfo{author}{H.~{Ferguson}},
  \bibinfo{author}{E.~{Figueroa}}, \bibinfo{author}{M.~{Fisher-Levine}},
  \bibinfo{author}{W.~{Focke}}, \bibinfo{author}{M.~D. {Foss}},
  \bibinfo{author}{J.~{Frank}}, \bibinfo{author}{M.~D. {Freemon}},
  \bibinfo{author}{E.~{Gangler}}, \bibinfo{author}{E.~{Gawiser}},
  \bibinfo{author}{J.~C. {Geary}}, \bibinfo{author}{P.~{Gee}},
  \bibinfo{author}{M.~{Geha}}, \bibinfo{author}{C.~J.~B. {Gessner}},
  \bibinfo{author}{R.~R. {Gibson}}, \bibinfo{author}{D.~K. {Gilmore}},
  \bibinfo{author}{T.~{Glanzman}}, \bibinfo{author}{W.~{Glick}},
  \bibinfo{author}{T.~{Goldina}}, \bibinfo{author}{D.~A. {Goldstein}},
  \bibinfo{author}{I.~{Goodenow}}, \bibinfo{author}{M.~L. {Graham}},
  \bibinfo{author}{W.~J. {Gressler}}, \bibinfo{author}{P.~{Gris}},
  \bibinfo{author}{L.~P. {Guy}}, \bibinfo{author}{A.~{Guyonnet}},
  \bibinfo{author}{G.~{Haller}}, \bibinfo{author}{R.~{Harris}},
  \bibinfo{author}{P.~A. {Hascall}}, \bibinfo{author}{J.~{Haupt}},
  \bibinfo{author}{F.~{Hernandez}}, \bibinfo{author}{S.~{Herrmann}},
  \bibinfo{author}{E.~{Hileman}}, \bibinfo{author}{J.~{Hoblitt}},
  \bibinfo{author}{J.~A. {Hodgson}}, \bibinfo{author}{C.~{Hogan}},
  \bibinfo{author}{J.~D. {Howard}}, \bibinfo{author}{D.~{Huang}},
  \bibinfo{author}{M.~E. {Huffer}}, \bibinfo{author}{P.~{Ingraham}},
  \bibinfo{author}{W.~R. {Innes}}, \bibinfo{author}{S.~H. {Jacoby}},
  \bibinfo{author}{B.~{Jain}}, \bibinfo{author}{F.~{Jammes}},
  \bibinfo{author}{M.~J. {Jee}}, \bibinfo{author}{T.~{Jenness}},
  \bibinfo{author}{G.~{Jernigan}}, \bibinfo{author}{D.~{Jevremovi{\'c}}},
  \bibinfo{author}{K.~{Johns}}, \bibinfo{author}{A.~S. {Johnson}},
  \bibinfo{author}{M.~W.~G. {Johnson}}, \bibinfo{author}{R.~L. {Jones}},
  \bibinfo{author}{C.~{Juramy-Gilles}}, \bibinfo{author}{M.~{Juri{\'c}}},
  \bibinfo{author}{J.~S. {Kalirai}}, \bibinfo{author}{N.~J. {Kallivayalil}},
  \bibinfo{author}{B.~{Kalmbach}}, \bibinfo{author}{J.~P. {Kantor}},
  \bibinfo{author}{P.~{Karst}}, \bibinfo{author}{M.~M. {Kasliwal}},
  \bibinfo{author}{H.~{Kelly}}, \bibinfo{author}{R.~{Kessler}},
  \bibinfo{author}{V.~{Kinnison}}, \bibinfo{author}{D.~{Kirkby}},
  \bibinfo{author}{L.~{Knox}}, \bibinfo{author}{I.~V. {Kotov}},
  \bibinfo{author}{V.~L. {Krabbendam}}, \bibinfo{author}{K.~S. {Krughoff}},
  \bibinfo{author}{P.~{Kub{\'a}nek}}, \bibinfo{author}{J.~{Kuczewski}},
  \bibinfo{author}{S.~{Kulkarni}}, \bibinfo{author}{J.~{Ku}},
  \bibinfo{author}{N.~R. {Kurita}}, \bibinfo{author}{C.~S. {Lage}},
  \bibinfo{author}{R.~{Lambert}}, \bibinfo{author}{T.~{Lange}},
  \bibinfo{author}{J.~B. {Langton}}, \bibinfo{author}{L.~{Le Guillou}},
  \bibinfo{author}{D.~{Levine}}, \bibinfo{author}{M.~{Liang}},
  \bibinfo{author}{K.-T. {Lim}}, \bibinfo{author}{C.~J. {Lintott}},
  \bibinfo{author}{K.~E. {Long}}, \bibinfo{author}{M.~{Lopez}},
  \bibinfo{author}{P.~J. {Lotz}}, \bibinfo{author}{R.~H. {Lupton}},
  \bibinfo{author}{N.~B. {Lust}}, \bibinfo{author}{L.~A. {MacArthur}},
  \bibinfo{author}{A.~{Mahabal}}, \bibinfo{author}{R.~{Mandelbaum}},
  \bibinfo{author}{T.~W. {Markiewicz}}, \bibinfo{author}{D.~S. {Marsh}},
  \bibinfo{author}{P.~J. {Marshall}}, \bibinfo{author}{S.~{Marshall}},
  \bibinfo{author}{M.~{May}}, \bibinfo{author}{R.~{McKercher}},
  \bibinfo{author}{M.~{McQueen}}, \bibinfo{author}{J.~{Meyers}},
  \bibinfo{author}{M.~{Migliore}}, \bibinfo{author}{M.~{Miller}},
  \bibinfo{author}{D.~J. {Mills}}, \bibinfo{author}{C.~{Miraval}},
  \bibinfo{author}{J.~{Moeyens}}, \bibinfo{author}{F.~E. {Moolekamp}},
  \bibinfo{author}{D.~G. {Monet}}, \bibinfo{author}{M.~{Moniez}},
  \bibinfo{author}{S.~{Monkewitz}}, \bibinfo{author}{C.~{Montgomery}},
  \bibinfo{author}{C.~B. {Morrison}}, \bibinfo{author}{F.~{Mueller}},
  \bibinfo{author}{G.~P. {Muller}}, \bibinfo{author}{F.~{Mu{\~n}oz Arancibia}},
  \bibinfo{author}{D.~R. {Neill}}, \bibinfo{author}{S.~P. {Newbry}},
  \bibinfo{author}{J.-Y. {Nief}}, \bibinfo{author}{A.~{Nomerotski}},
  \bibinfo{author}{M.~{Nordby}}, \bibinfo{author}{P.~{O'Connor}},
  \bibinfo{author}{J.~{Oliver}}, \bibinfo{author}{S.~S. {Olivier}},
  \bibinfo{author}{K.~{Olsen}}, \bibinfo{author}{W.~{O'Mullane}},
  \bibinfo{author}{S.~{Ortiz}}, \bibinfo{author}{S.~{Osier}},
  \bibinfo{author}{R.~E. {Owen}}, \bibinfo{author}{R.~{Pain}},
  \bibinfo{author}{P.~E. {Palecek}}, \bibinfo{author}{J.~K. {Parejko}},
  \bibinfo{author}{J.~B. {Parsons}}, \bibinfo{author}{N.~M. {Pease}},
  \bibinfo{author}{J.~M. {Peterson}}, \bibinfo{author}{J.~R. {Peterson}},
  \bibinfo{author}{D.~L. {Petravick}}, \bibinfo{author}{M.~E. {Libby Petrick}},
  \bibinfo{author}{C.~E. {Petry}}, \bibinfo{author}{F.~{Pierfederici}},
  \bibinfo{author}{S.~{Pietrowicz}}, \bibinfo{author}{R.~{Pike}},
  \bibinfo{author}{P.~A. {Pinto}}, \bibinfo{author}{R.~{Plante}},
  \bibinfo{author}{S.~{Plate}}, \bibinfo{author}{J.~P. {Plutchak}},
  \bibinfo{author}{P.~A. {Price}}, \bibinfo{author}{M.~{Prouza}},
  \bibinfo{author}{V.~{Radeka}}, \bibinfo{author}{J.~{Rajagopal}},
  \bibinfo{author}{A.~P. {Rasmussen}}, \bibinfo{author}{N.~{Regnault}},
  \bibinfo{author}{K.~A. {Reil}}, \bibinfo{author}{D.~J. {Reiss}},
  \bibinfo{author}{M.~A. {Reuter}}, \bibinfo{author}{S.~T. {Ridgway}},
  \bibinfo{author}{V.~J. {Riot}}, \bibinfo{author}{S.~{Ritz}},
  \bibinfo{author}{S.~{Robinson}}, \bibinfo{author}{W.~{Roby}},
  \bibinfo{author}{A.~{Roodman}}, \bibinfo{author}{W.~{Rosing}},
  \bibinfo{author}{C.~{Roucelle}}, \bibinfo{author}{M.~R. {Rumore}},
  \bibinfo{author}{S.~{Russo}}, \bibinfo{author}{A.~{Saha}},
  \bibinfo{author}{B.~{Sassolas}}, \bibinfo{author}{T.~L. {Schalk}},
  \bibinfo{author}{P.~{Schellart}}, \bibinfo{author}{R.~H. {Schindler}},
  \bibinfo{author}{S.~{Schmidt}}, \bibinfo{author}{D.~P. {Schneider}},
  \bibinfo{author}{M.~D. {Schneider}}, \bibinfo{author}{W.~{Schoening}},
  \bibinfo{author}{G.~{Schumacher}}, \bibinfo{author}{M.~E. {Schwamb}},
  \bibinfo{author}{J.~{Sebag}}, \bibinfo{author}{B.~{Selvy}},
  \bibinfo{author}{G.~H. {Sembroski}}, \bibinfo{author}{L.~G. {Seppala}},
  \bibinfo{author}{A.~{Serio}}, \bibinfo{author}{E.~{Serrano}},
  \bibinfo{author}{R.~A. {Shaw}}, \bibinfo{author}{I.~{Shipsey}},
  \bibinfo{author}{J.~{Sick}}, \bibinfo{author}{N.~{Silvestri}},
  \bibinfo{author}{C.~T. {Slater}}, \bibinfo{author}{J.~A. {Smith}},
  \bibinfo{author}{R.~C. {Smith}}, \bibinfo{author}{S.~{Sobhani}},
  \bibinfo{author}{C.~{Soldahl}}, \bibinfo{author}{L.~{Storrie-Lombardi}},
  \bibinfo{author}{E.~{Stover}}, \bibinfo{author}{M.~A. {Strauss}},
  \bibinfo{author}{R.~A. {Street}}, \bibinfo{author}{C.~W. {Stubbs}},
  \bibinfo{author}{I.~S. {Sullivan}}, \bibinfo{author}{D.~{Sweeney}},
  \bibinfo{author}{J.~D. {Swinbank}}, \bibinfo{author}{A.~{Szalay}},
  \bibinfo{author}{P.~{Takacs}}, \bibinfo{author}{S.~A. {Tether}},
  \bibinfo{author}{J.~J. {Thaler}}, \bibinfo{author}{J.~G. {Thayer}},
  \bibinfo{author}{S.~{Thomas}}, \bibinfo{author}{A.~J. {Thornton}},
  \bibinfo{author}{V.~{Thukral}}, \bibinfo{author}{J.~{Tice}},
  \bibinfo{author}{D.~E. {Trilling}}, \bibinfo{author}{M.~{Turri}},
  \bibinfo{author}{R.~{Van Berg}}, \bibinfo{author}{D.~{Vanden Berk}},
  \bibinfo{author}{K.~{Vetter}}, \bibinfo{author}{F.~{Virieux}},
  \bibinfo{author}{T.~{Vucina}}, \bibinfo{author}{W.~{Wahl}},
  \bibinfo{author}{L.~{Walkowicz}}, \bibinfo{author}{B.~{Walsh}},
  \bibinfo{author}{C.~W. {Walter}}, \bibinfo{author}{D.~L. {Wang}},
  \bibinfo{author}{S.-Y. {Wang}}, \bibinfo{author}{M.~{Warner}},
  \bibinfo{author}{O.~{Wiecha}}, \bibinfo{author}{B.~{Willman}},
  \bibinfo{author}{S.~E. {Winters}}, \bibinfo{author}{D.~{Wittman}},
  \bibinfo{author}{S.~C. {Wolff}}, \bibinfo{author}{W.~M. {Wood-Vasey}},
  \bibinfo{author}{X.~{Wu}}, \bibinfo{author}{B.~{Xin}},
  \bibinfo{author}{P.~{Yoachim}}, \bibinfo{author}{H.~{Zhan}},
\newblock \bibinfo{title}{{LSST: From Science Drivers to Reference Design and
  Anticipated Data Products}},
\newblock \bibinfo{journal}{ApJ} \bibinfo{volume}{873} (\bibinfo{year}{2019})
  \bibinfo{pages}{111}.
\bibitem[{{Bellm} et~al.(2019){Bellm}, {Kulkarni}, {Graham}, {Dekany}, {Smith},
  {Riddle}, {Masci}, {Helou}, {Prince}, {Adams}, {Barbarino}, {Barlow},
  {Bauer}, {Beck}, {Belicki}, {Biswas}, {Blagorodnova}, {Bodewits}, {Bolin},
  {Brinnel}, {Brooke}, {Bue}, {Bulla}, {Burruss}, {Cenko}, {Chang}, {Connolly},
  {Coughlin}, {Cromer}, {Cunningham}, {De}, {Delacroix}, {Desai}, {Duev},
  {Eadie}, {Farnham}, {Feeney}, {Feindt}, {Flynn}, {Franckowiak}, {Frederick},
  {Fremling}, {Gal-Yam}, {Gezari}, {Giomi}, {Goldstein}, {Golkhou}, {Goobar},
  {Groom}, {Hacopians}, {Hale}, {Henning}, {Ho}, {Hover}, {Howell}, {Hung},
  {Huppenkothen}, {Imel}, {Ip}, {Ivezi{\'c}}, {Jackson}, {Jones}, {Juric},
  {Kasliwal}, {Kaspi}, {Kaye}, {Kelley}, {Kowalski}, {Kramer}, {Kupfer},
  {Landry}, {Laher}, {Lee}, {Lin}, {Lin}, {Lunnan}, {Giomi}, {Mahabal}, {Mao},
  {Miller}, {Monkewitz}, {Murphy}, {Ngeow}, {Nordin}, {Nugent}, {Ofek},
  {Patterson}, {Penprase}, {Porter}, {Rauch}, {Rebbapragada}, {Reiley},
  {Rigault}, {Rodriguez}, {van Roestel}, {Rusholme}, {van Santen}, {Schulze},
  {Shupe}, {Singer}, {Soumagnac}, {Stein}, {Surace}, {Sollerman}, {Szkody},
  {Taddia}, {Terek}, {Van Sistine}, {van Velzen}, {Vestrand}, {Walters},
  {Ward}, {Ye}, {Yu}, {Yan}, and {Zolkower}}]{2019PASP..131a8002B}
\bibinfo{author}{E.~C. {Bellm}}, \bibinfo{author}{S.~R. {Kulkarni}},
  \bibinfo{author}{M.~J. {Graham}}, \bibinfo{author}{R.~{Dekany}},
  \bibinfo{author}{R.~M. {Smith}}, \bibinfo{author}{R.~{Riddle}},
  \bibinfo{author}{F.~J. {Masci}}, \bibinfo{author}{G.~{Helou}},
  \bibinfo{author}{T.~A. {Prince}}, \bibinfo{author}{S.~M. {Adams}},
  \bibinfo{author}{C.~{Barbarino}}, \bibinfo{author}{T.~{Barlow}},
  \bibinfo{author}{J.~{Bauer}}, \bibinfo{author}{R.~{Beck}},
  \bibinfo{author}{J.~{Belicki}}, \bibinfo{author}{R.~{Biswas}},
  \bibinfo{author}{N.~{Blagorodnova}}, \bibinfo{author}{D.~{Bodewits}},
  \bibinfo{author}{B.~{Bolin}}, \bibinfo{author}{V.~{Brinnel}},
  \bibinfo{author}{T.~{Brooke}}, \bibinfo{author}{B.~{Bue}},
  \bibinfo{author}{M.~{Bulla}}, \bibinfo{author}{R.~{Burruss}},
  \bibinfo{author}{S.~B. {Cenko}}, \bibinfo{author}{C.-K. {Chang}},
  \bibinfo{author}{A.~{Connolly}}, \bibinfo{author}{M.~{Coughlin}},
  \bibinfo{author}{J.~{Cromer}}, \bibinfo{author}{V.~{Cunningham}},
  \bibinfo{author}{K.~{De}}, \bibinfo{author}{A.~{Delacroix}},
  \bibinfo{author}{V.~{Desai}}, \bibinfo{author}{D.~A. {Duev}},
  \bibinfo{author}{G.~{Eadie}}, \bibinfo{author}{T.~L. {Farnham}},
  \bibinfo{author}{M.~{Feeney}}, \bibinfo{author}{U.~{Feindt}},
  \bibinfo{author}{D.~{Flynn}}, \bibinfo{author}{A.~{Franckowiak}},
  \bibinfo{author}{S.~{Frederick}}, \bibinfo{author}{C.~{Fremling}},
  \bibinfo{author}{A.~{Gal-Yam}}, \bibinfo{author}{S.~{Gezari}},
  \bibinfo{author}{M.~{Giomi}}, \bibinfo{author}{D.~A. {Goldstein}},
  \bibinfo{author}{V.~Z. {Golkhou}}, \bibinfo{author}{A.~{Goobar}},
  \bibinfo{author}{S.~{Groom}}, \bibinfo{author}{E.~{Hacopians}},
  \bibinfo{author}{D.~{Hale}}, \bibinfo{author}{J.~{Henning}},
  \bibinfo{author}{A.~Y.~Q. {Ho}}, \bibinfo{author}{D.~{Hover}},
  \bibinfo{author}{J.~{Howell}}, \bibinfo{author}{T.~{Hung}},
  \bibinfo{author}{D.~{Huppenkothen}}, \bibinfo{author}{D.~{Imel}},
  \bibinfo{author}{W.-H. {Ip}}, \bibinfo{author}{{\v{Z}}.~{Ivezi{\'c}}},
  \bibinfo{author}{E.~{Jackson}}, \bibinfo{author}{L.~{Jones}},
  \bibinfo{author}{M.~{Juric}}, \bibinfo{author}{M.~M. {Kasliwal}},
  \bibinfo{author}{S.~{Kaspi}}, \bibinfo{author}{S.~{Kaye}},
  \bibinfo{author}{M.~S.~P. {Kelley}}, \bibinfo{author}{M.~{Kowalski}},
  \bibinfo{author}{E.~{Kramer}}, \bibinfo{author}{T.~{Kupfer}},
  \bibinfo{author}{W.~{Landry}}, \bibinfo{author}{R.~R. {Laher}},
  \bibinfo{author}{C.-D. {Lee}}, \bibinfo{author}{H.~W. {Lin}},
  \bibinfo{author}{Z.-Y. {Lin}}, \bibinfo{author}{R.~{Lunnan}},
  \bibinfo{author}{M.~{Giomi}}, \bibinfo{author}{A.~{Mahabal}},
  \bibinfo{author}{P.~{Mao}}, \bibinfo{author}{A.~A. {Miller}},
  \bibinfo{author}{S.~{Monkewitz}}, \bibinfo{author}{P.~{Murphy}},
  \bibinfo{author}{C.-C. {Ngeow}}, \bibinfo{author}{J.~{Nordin}},
  \bibinfo{author}{P.~{Nugent}}, \bibinfo{author}{E.~{Ofek}},
  \bibinfo{author}{M.~T. {Patterson}}, \bibinfo{author}{B.~{Penprase}},
  \bibinfo{author}{M.~{Porter}}, \bibinfo{author}{L.~{Rauch}},
  \bibinfo{author}{U.~{Rebbapragada}}, \bibinfo{author}{D.~{Reiley}},
  \bibinfo{author}{M.~{Rigault}}, \bibinfo{author}{H.~{Rodriguez}},
  \bibinfo{author}{J.~{van Roestel}}, \bibinfo{author}{B.~{Rusholme}},
  \bibinfo{author}{J.~{van Santen}}, \bibinfo{author}{S.~{Schulze}},
  \bibinfo{author}{D.~L. {Shupe}}, \bibinfo{author}{L.~P. {Singer}},
  \bibinfo{author}{M.~T. {Soumagnac}}, \bibinfo{author}{R.~{Stein}},
  \bibinfo{author}{J.~{Surace}}, \bibinfo{author}{J.~{Sollerman}},
  \bibinfo{author}{P.~{Szkody}}, \bibinfo{author}{F.~{Taddia}},
  \bibinfo{author}{S.~{Terek}}, \bibinfo{author}{A.~{Van Sistine}},
  \bibinfo{author}{S.~{van Velzen}}, \bibinfo{author}{W.~T. {Vestrand}},
  \bibinfo{author}{R.~{Walters}}, \bibinfo{author}{C.~{Ward}},
  \bibinfo{author}{Q.-Z. {Ye}}, \bibinfo{author}{P.-C. {Yu}},
  \bibinfo{author}{L.~{Yan}}, \bibinfo{author}{J.~{Zolkower}},
\newblock \bibinfo{title}{{The Zwicky Transient Facility: System Overview,
  Performance, and First Results}},
\newblock \bibinfo{journal}{PASP} \bibinfo{volume}{131} (\bibinfo{year}{2019})
  \bibinfo{pages}{018002}.
\bibitem[{{Jonas} and {MeerKAT Team}(2016)}]{2016mks..confE...1J}
\bibinfo{author}{J.~{Jonas}}, \bibinfo{author}{{MeerKAT Team}},
\newblock \bibinfo{title}{{The MeerKAT Radio Telescope}},
\newblock in: \bibinfo{booktitle}{Proceedings of MeerKAT Science: On the
  Pathway to the SKA. 25-27 May}, \bibinfo{year}{2016}, p.~\bibinfo{pages}{1}.
\bibitem[{{Murphy} et~al.(2021){Murphy}, {Kaplan}, {Stewart}, {O'Brien},
  {Lenc}, {Pintaldi}, {Pritchard}, {Dobie}, {Fox}, {Leung}, {An}, {Bell},
  {Broderick}, {Chatterjee}, {Dai}, {d'Antonio}, {Doyle}, {Gaensler}, {Heald},
  {Horesh}, {Jones}, {McConnell}, {Moss}, {Raja}, {Ramsay}, {Ryder}, {Sadler},
  {Sivakoff}, {Wang}, {Wang}, {Wheatland}, {Whiting}, {Allison}, {Anderson},
  {Ball}, {Bannister}, {Bock}, {Bolton}, {Bunton}, {Chekkala}, {Chippendale},
  {Cooray}, {Gupta}, {Hayman}, {Jeganathan}, {Koribalski}, {Lee-Waddell},
  {Mahony}, {Marvil}, {McClure-Griffiths}, {Mirtschin}, {Ng}, {Pearce},
  {Phillips}, and {Voronkov}}]{2021PASA...38...54M}
\bibinfo{author}{T.~{Murphy}}, \bibinfo{author}{D.~L. {Kaplan}},
  \bibinfo{author}{A.~J. {Stewart}}, \bibinfo{author}{A.~{O'Brien}},
  \bibinfo{author}{E.~{Lenc}}, \bibinfo{author}{S.~{Pintaldi}},
  \bibinfo{author}{J.~{Pritchard}}, \bibinfo{author}{D.~{Dobie}},
  \bibinfo{author}{A.~{Fox}}, \bibinfo{author}{J.~K. {Leung}},
  \bibinfo{author}{T.~{An}}, \bibinfo{author}{M.~E. {Bell}},
  \bibinfo{author}{J.~W. {Broderick}}, \bibinfo{author}{S.~{Chatterjee}},
  \bibinfo{author}{S.~{Dai}}, \bibinfo{author}{D.~{d'Antonio}},
  \bibinfo{author}{G.~{Doyle}}, \bibinfo{author}{B.~M. {Gaensler}},
  \bibinfo{author}{G.~{Heald}}, \bibinfo{author}{A.~{Horesh}},
  \bibinfo{author}{M.~L. {Jones}}, \bibinfo{author}{D.~{McConnell}},
  \bibinfo{author}{V.~A. {Moss}}, \bibinfo{author}{W.~{Raja}},
  \bibinfo{author}{G.~{Ramsay}}, \bibinfo{author}{S.~{Ryder}},
  \bibinfo{author}{E.~M. {Sadler}}, \bibinfo{author}{G.~R. {Sivakoff}},
  \bibinfo{author}{Y.~{Wang}}, \bibinfo{author}{Z.~{Wang}},
  \bibinfo{author}{M.~S. {Wheatland}}, \bibinfo{author}{M.~{Whiting}},
  \bibinfo{author}{J.~R. {Allison}}, \bibinfo{author}{C.~S. {Anderson}},
  \bibinfo{author}{L.~{Ball}}, \bibinfo{author}{K.~{Bannister}},
  \bibinfo{author}{D.~C.~J. {Bock}}, \bibinfo{author}{R.~{Bolton}},
  \bibinfo{author}{J.~D. {Bunton}}, \bibinfo{author}{R.~{Chekkala}},
  \bibinfo{author}{A.~P. {Chippendale}}, \bibinfo{author}{F.~R. {Cooray}},
  \bibinfo{author}{N.~{Gupta}}, \bibinfo{author}{D.~B. {Hayman}},
  \bibinfo{author}{K.~{Jeganathan}}, \bibinfo{author}{B.~{Koribalski}},
  \bibinfo{author}{K.~{Lee-Waddell}}, \bibinfo{author}{E.~K. {Mahony}},
  \bibinfo{author}{J.~{Marvil}}, \bibinfo{author}{N.~M. {McClure-Griffiths}},
  \bibinfo{author}{P.~{Mirtschin}}, \bibinfo{author}{A.~{Ng}},
  \bibinfo{author}{S.~{Pearce}}, \bibinfo{author}{C.~{Phillips}},
  \bibinfo{author}{M.~A. {Voronkov}},
\newblock \bibinfo{title}{{The ASKAP Variables and Slow Transients (VAST) Pilot
  Survey}},
\newblock \bibinfo{journal}{PASA} \bibinfo{volume}{38} (\bibinfo{year}{2021})
  \bibinfo{pages}{e054}.
\bibitem[{{CHIME/FRB Collaboration} et~al.(2019){CHIME/FRB Collaboration},
  {Amiri}, {Bandura}, {Bhardwaj}, {Boubel}, {Boyce}, {Boyle}, {. Brar},
  {Burhanpurkar}, {Cassanelli}, {Chawla}, {Cliche}, {Cubranic}, {Deng},
  {Denman}, {Dobbs}, {Fandino}, {Fonseca}, {Gaensler}, {Gilbert}, {Gill},
  {Giri}, {Good}, {Halpern}, {Hanna}, {Hill}, {Hinshaw}, {H{\"o}fer},
  {Josephy}, {Kaspi}, {Landecker}, {Lang}, {Lin}, {Masui}, {Mckinven},
  {Mena-Parra}, {Merryfield}, {Michilli}, {Milutinovic}, {Moatti}, {Naidu},
  {Newburgh}, {Ng}, {Patel}, {Pen}, {Pinsonneault-Marotte}, {Pleunis},
  {Rafiei-Ravandi}, {Rahman}, {Ransom}, {Renard}, {Scholz}, {Shaw}, {Siegel},
  {Smith}, {Stairs}, {Tendulkar}, {Tretyakov}, {Vanderlinde}, and
  {Yadav}}]{2019Natur.566..235C}
\bibinfo{author}{{CHIME/FRB Collaboration}}, \bibinfo{author}{M.~{Amiri}},
  \bibinfo{author}{K.~{Bandura}}, \bibinfo{author}{M.~{Bhardwaj}},
  \bibinfo{author}{P.~{Boubel}}, \bibinfo{author}{M.~M. {Boyce}},
  \bibinfo{author}{P.~J. {Boyle}}, \bibinfo{author}{C.~{. Brar}},
  \bibinfo{author}{M.~{Burhanpurkar}}, \bibinfo{author}{T.~{Cassanelli}},
  \bibinfo{author}{P.~{Chawla}}, \bibinfo{author}{J.~F. {Cliche}},
  \bibinfo{author}{D.~{Cubranic}}, \bibinfo{author}{M.~{Deng}},
  \bibinfo{author}{N.~{Denman}}, \bibinfo{author}{M.~{Dobbs}},
  \bibinfo{author}{M.~{Fandino}}, \bibinfo{author}{E.~{Fonseca}},
  \bibinfo{author}{B.~M. {Gaensler}}, \bibinfo{author}{A.~J. {Gilbert}},
  \bibinfo{author}{A.~{Gill}}, \bibinfo{author}{U.~{Giri}},
  \bibinfo{author}{D.~C. {Good}}, \bibinfo{author}{M.~{Halpern}},
  \bibinfo{author}{D.~S. {Hanna}}, \bibinfo{author}{A.~S. {Hill}},
  \bibinfo{author}{G.~{Hinshaw}}, \bibinfo{author}{C.~{H{\"o}fer}},
  \bibinfo{author}{A.~{Josephy}}, \bibinfo{author}{V.~M. {Kaspi}},
  \bibinfo{author}{T.~L. {Landecker}}, \bibinfo{author}{D.~A. {Lang}},
  \bibinfo{author}{H.~H. {Lin}}, \bibinfo{author}{K.~W. {Masui}},
  \bibinfo{author}{R.~{Mckinven}}, \bibinfo{author}{J.~{Mena-Parra}},
  \bibinfo{author}{M.~{Merryfield}}, \bibinfo{author}{D.~{Michilli}},
  \bibinfo{author}{N.~{Milutinovic}}, \bibinfo{author}{C.~{Moatti}},
  \bibinfo{author}{A.~{Naidu}}, \bibinfo{author}{L.~B. {Newburgh}},
  \bibinfo{author}{C.~{Ng}}, \bibinfo{author}{C.~{Patel}},
  \bibinfo{author}{U.~{Pen}}, \bibinfo{author}{T.~{Pinsonneault-Marotte}},
  \bibinfo{author}{Z.~{Pleunis}}, \bibinfo{author}{M.~{Rafiei-Ravandi}},
  \bibinfo{author}{M.~{Rahman}}, \bibinfo{author}{S.~M. {Ransom}},
  \bibinfo{author}{A.~{Renard}}, \bibinfo{author}{P.~{Scholz}},
  \bibinfo{author}{J.~R. {Shaw}}, \bibinfo{author}{S.~R. {Siegel}},
  \bibinfo{author}{K.~M. {Smith}}, \bibinfo{author}{I.~H. {Stairs}},
  \bibinfo{author}{S.~P. {Tendulkar}}, \bibinfo{author}{I.~{Tretyakov}},
  \bibinfo{author}{K.~{Vanderlinde}}, \bibinfo{author}{P.~{Yadav}},
\newblock \bibinfo{title}{{A second source of repeating fast radio bursts}},
\newblock \bibinfo{journal}{Nature} \bibinfo{volume}{566}
  (\bibinfo{year}{2019}) \bibinfo{pages}{235--238}.
\bibitem[{Abbott et~al.(2017)}]{2017PhRvL.119p1101A}
\bibinfo{author}{R.~P. Abbott}, et~al.,
\newblock \bibinfo{title}{{GW170817: Observation of Gravitational Waves from a
  Binary Neutron Star Inspiral}},
\newblock \bibinfo{journal}{Physical Review Letters} \bibinfo{volume}{119}
  (\bibinfo{year}{2017}) \bibinfo{pages}{161101}.
\bibitem[{{Troja} et~al.(2017){Troja}, {Piro}, {van Eerten}, {Wollaeger}, {Im},
  {Fox}, {Butler}, {Cenko}, {Sakamoto}, {Fryer}, {Ricci}, {Lien}, {Ryan},
  {Korobkin}, {Lee}, {Burgess}, {Lee}, {Watson}, {Choi}, {Covino}, {D'Avanzo},
  {Fontes}, {González}, {Khandrika}, {Kim}, {Kim}, {Lee}, {Lee}, {Kutyrev},
  {Lim}, {Sánchez-Ramírez}, {Veilleux}, {Wieringa}, and
  {Yoon}}]{2017Natur.551...71T}
\bibinfo{author}{E.~{Troja}}, \bibinfo{author}{L.~{Piro}},
  \bibinfo{author}{H.~{van Eerten}}, \bibinfo{author}{R.~T. {Wollaeger}},
  \bibinfo{author}{M.~{Im}}, \bibinfo{author}{O.~D. {Fox}},
  \bibinfo{author}{N.~R. {Butler}}, \bibinfo{author}{S.~B. {Cenko}},
  \bibinfo{author}{T.~{Sakamoto}}, \bibinfo{author}{C.~L. {Fryer}},
  \bibinfo{author}{R.~{Ricci}}, \bibinfo{author}{A.~{Lien}},
  \bibinfo{author}{R.~E. {Ryan}}, \bibinfo{author}{O.~{Korobkin}},
  \bibinfo{author}{S.~K. {Lee}}, \bibinfo{author}{J.~M. {Burgess}},
  \bibinfo{author}{W.~H. {Lee}}, \bibinfo{author}{A.~M. {Watson}},
  \bibinfo{author}{C.~{Choi}}, \bibinfo{author}{S.~{Covino}},
  \bibinfo{author}{P.~{D'Avanzo}}, \bibinfo{author}{C.~J. {Fontes}},
  \bibinfo{author}{J.~B. {González}}, \bibinfo{author}{H.~G. {Khandrika}},
  \bibinfo{author}{J.~{Kim}}, \bibinfo{author}{S.~L. {Kim}},
  \bibinfo{author}{C.~U. {Lee}}, \bibinfo{author}{H.~M. {Lee}},
  \bibinfo{author}{A.~{Kutyrev}}, \bibinfo{author}{G.~{Lim}},
  \bibinfo{author}{R.~{Sánchez-Ramírez}}, \bibinfo{author}{S.~{Veilleux}},
  \bibinfo{author}{M.~H. {Wieringa}}, \bibinfo{author}{Y.~{Yoon}},
\newblock \bibinfo{title}{{The X-ray counterpart to the gravitational-wave
  event GW170817}},
\newblock \bibinfo{journal}{Nature} \bibinfo{volume}{551}
  (\bibinfo{year}{2017}) \bibinfo{pages}{71--74}.
\bibitem[{{Mooley} et~al.(2018){Mooley}, {Frail}, {Dobie}, {Lenc}, {Corsi},
  {De}, {Nayana}, {Makhathini}, {Heywood}, {Murphy}, {Kaplan}, {Chandra},
  {Smirnov}, {Nakar}, {Hallinan}, {Camilo}, {Fender}, {Goedhart}, {Groot},
  {Kasliwal}, {Kulkarni}, and {Woudt}}]{2018ApJ...868L..11M}
\bibinfo{author}{K.~P. {Mooley}}, \bibinfo{author}{D.~A. {Frail}},
  \bibinfo{author}{D.~{Dobie}}, \bibinfo{author}{E.~{Lenc}},
  \bibinfo{author}{A.~{Corsi}}, \bibinfo{author}{K.~{De}},
  \bibinfo{author}{A.~J. {Nayana}}, \bibinfo{author}{S.~{Makhathini}},
  \bibinfo{author}{I.~{Heywood}}, \bibinfo{author}{T.~{Murphy}},
  \bibinfo{author}{D.~L. {Kaplan}}, \bibinfo{author}{P.~{Chandra}},
  \bibinfo{author}{O.~{Smirnov}}, \bibinfo{author}{E.~{Nakar}},
  \bibinfo{author}{G.~{Hallinan}}, \bibinfo{author}{F.~{Camilo}},
  \bibinfo{author}{R.~{Fender}}, \bibinfo{author}{S.~{Goedhart}},
  \bibinfo{author}{P.~{Groot}}, \bibinfo{author}{M.~M. {Kasliwal}},
  \bibinfo{author}{S.~R. {Kulkarni}}, \bibinfo{author}{P.~A. {Woudt}},
\newblock \bibinfo{title}{{A Strong Jet Signature in the Late-time Light Curve
  of GW170817}},
\newblock \bibinfo{journal}{ApJL} \bibinfo{volume}{868} (\bibinfo{year}{2018})
  \bibinfo{pages}{L11}.
\bibitem[{{Frail} et~al.(1997){Frail}, {Kulkarni}, {Nicastro}, {Feroci}, and
  {Taylor}}]{1997Natur.389..261F}
\bibinfo{author}{D.~A. {Frail}}, \bibinfo{author}{S.~R. {Kulkarni}},
  \bibinfo{author}{L.~{Nicastro}}, \bibinfo{author}{M.~{Feroci}},
  \bibinfo{author}{G.~B. {Taylor}},
\newblock \bibinfo{title}{{The radio afterglow from the
  {\ensuremath{\gamma}}-ray burst of 8 May 1997}},
\newblock \bibinfo{journal}{Nature} \bibinfo{volume}{389}
  (\bibinfo{year}{1997}) \bibinfo{pages}{261--263}.
\bibitem[{{Anderson} et~al.(2018){Anderson}, {Staley}, {van der Horst},
  {Fender}, {Rowlinson}, {Mooley}, {Broderick}, {Wijers}, {Rumsey}, and
  {Titterington}}]{2018MNRAS.473.1512A}
\bibinfo{author}{G.~E. {Anderson}}, \bibinfo{author}{T.~D. {Staley}},
  \bibinfo{author}{A.~J. {van der Horst}}, \bibinfo{author}{R.~P. {Fender}},
  \bibinfo{author}{A.~{Rowlinson}}, \bibinfo{author}{K.~P. {Mooley}},
  \bibinfo{author}{J.~W. {Broderick}}, \bibinfo{author}{R.~A.~M.~J. {Wijers}},
  \bibinfo{author}{C.~{Rumsey}}, \bibinfo{author}{D.~J. {Titterington}},
\newblock \bibinfo{title}{{The Arcminute Microkelvin Imager catalogue of
  gamma-ray burst afterglows at 15.7 GHz}},
\newblock \bibinfo{journal}{MNRAS} \bibinfo{volume}{473} (\bibinfo{year}{2018})
  \bibinfo{pages}{1512--1536}.
\bibitem[{{Levan} et~al.(2011){Levan}, {Tanvir}, {Cenko}, {Perley}, {Wiersema},
  {Bloom}, {Fruchter}, {de Ugarte Postigo}, {O'Brien}, {Butler}, {van der
  Horst}, {Leloudas}, {Morgan}, {Misra}, {Bower}, {Farihi}, {Tunnicliffe},
  {Modjaz}, {Silverman}, {Hjorth}, {Th{\"o}ne}, {Cucchiara}, {Cer{\'o}n},
  {Castro-Tirado}, {Arnold}, {Bremer}, {Brodie}, {Carroll}, {Cooper}, {Curran},
  {Cutri}, {Ehle}, {Forbes}, {Fynbo}, {Gorosabel}, {Graham}, {Hoffman},
  {Guziy}, {Jakobsson}, {Kamble}, {Kerr}, {Kasliwal}, {Kouveliotou},
  {Kocevski}, {Law}, {Nugent}, {Ofek}, {Poznanski}, {Quimby}, {Rol},
  {Romanowsky}, {S{\'a}nchez-Ram{\'\i}rez}, {Schulze}, {Singh}, {van
  Spaandonk}, {Starling}, {Strom}, {Tello}, {Vaduvescu}, {Wheatley}, {Wijers},
  {Winters}, and {Xu}}]{2011Sci...333..199L}
\bibinfo{author}{A.~J. {Levan}}, \bibinfo{author}{N.~R. {Tanvir}},
  \bibinfo{author}{S.~B. {Cenko}}, \bibinfo{author}{D.~A. {Perley}},
  \bibinfo{author}{K.~{Wiersema}}, \bibinfo{author}{J.~S. {Bloom}},
  \bibinfo{author}{A.~S. {Fruchter}}, \bibinfo{author}{A.~{de Ugarte Postigo}},
  \bibinfo{author}{P.~T. {O'Brien}}, \bibinfo{author}{N.~{Butler}},
  \bibinfo{author}{A.~J. {van der Horst}}, \bibinfo{author}{G.~{Leloudas}},
  \bibinfo{author}{A.~N. {Morgan}}, \bibinfo{author}{K.~{Misra}},
  \bibinfo{author}{G.~C. {Bower}}, \bibinfo{author}{J.~{Farihi}},
  \bibinfo{author}{R.~L. {Tunnicliffe}}, \bibinfo{author}{M.~{Modjaz}},
  \bibinfo{author}{J.~M. {Silverman}}, \bibinfo{author}{J.~{Hjorth}},
  \bibinfo{author}{C.~{Th{\"o}ne}}, \bibinfo{author}{A.~{Cucchiara}},
  \bibinfo{author}{J.~M.~C. {Cer{\'o}n}}, \bibinfo{author}{A.~J.
  {Castro-Tirado}}, \bibinfo{author}{J.~A. {Arnold}},
  \bibinfo{author}{M.~{Bremer}}, \bibinfo{author}{J.~P. {Brodie}},
  \bibinfo{author}{T.~{Carroll}}, \bibinfo{author}{M.~C. {Cooper}},
  \bibinfo{author}{P.~A. {Curran}}, \bibinfo{author}{R.~M. {Cutri}},
  \bibinfo{author}{J.~{Ehle}}, \bibinfo{author}{D.~{Forbes}},
  \bibinfo{author}{J.~{Fynbo}}, \bibinfo{author}{J.~{Gorosabel}},
  \bibinfo{author}{J.~{Graham}}, \bibinfo{author}{D.~I. {Hoffman}},
  \bibinfo{author}{S.~{Guziy}}, \bibinfo{author}{P.~{Jakobsson}},
  \bibinfo{author}{A.~{Kamble}}, \bibinfo{author}{T.~{Kerr}},
  \bibinfo{author}{M.~M. {Kasliwal}}, \bibinfo{author}{C.~{Kouveliotou}},
  \bibinfo{author}{D.~{Kocevski}}, \bibinfo{author}{N.~M. {Law}},
  \bibinfo{author}{P.~E. {Nugent}}, \bibinfo{author}{E.~O. {Ofek}},
  \bibinfo{author}{D.~{Poznanski}}, \bibinfo{author}{R.~M. {Quimby}},
  \bibinfo{author}{E.~{Rol}}, \bibinfo{author}{A.~J. {Romanowsky}},
  \bibinfo{author}{R.~{S{\'a}nchez-Ram{\'\i}rez}},
  \bibinfo{author}{S.~{Schulze}}, \bibinfo{author}{N.~{Singh}},
  \bibinfo{author}{L.~{van Spaandonk}}, \bibinfo{author}{R.~L.~C. {Starling}},
  \bibinfo{author}{R.~G. {Strom}}, \bibinfo{author}{J.~C. {Tello}},
  \bibinfo{author}{O.~{Vaduvescu}}, \bibinfo{author}{P.~J. {Wheatley}},
  \bibinfo{author}{R.~A.~M.~J. {Wijers}}, \bibinfo{author}{J.~M. {Winters}},
  \bibinfo{author}{D.~{Xu}},
\newblock \bibinfo{title}{{An Extremely Luminous Panchromatic Outburst from the
  Nucleus of a Distant Galaxy}},
\newblock \bibinfo{journal}{Science} \bibinfo{volume}{333}
  (\bibinfo{year}{2011}) \bibinfo{pages}{199}.
\bibitem[{{van Velzen} et~al.(2016){van Velzen}, {Anderson}, {Stone}, {Fraser},
  {Wevers}, {Metzger}, {Jonker}, {van der Horst}, {Staley}, {Mendez},
  {Miller-Jones}, {Hodgkin}, {Campbell}, and {Fender}}]{2016Sci...351...62V}
\bibinfo{author}{S.~{van Velzen}}, \bibinfo{author}{G.~E. {Anderson}},
  \bibinfo{author}{N.~C. {Stone}}, \bibinfo{author}{M.~{Fraser}},
  \bibinfo{author}{T.~{Wevers}}, \bibinfo{author}{B.~D. {Metzger}},
  \bibinfo{author}{P.~G. {Jonker}}, \bibinfo{author}{A.~J. {van der Horst}},
  \bibinfo{author}{T.~D. {Staley}}, \bibinfo{author}{A.~J. {Mendez}},
  \bibinfo{author}{J.~C.~A. {Miller-Jones}}, \bibinfo{author}{S.~T. {Hodgkin}},
  \bibinfo{author}{H.~C. {Campbell}}, \bibinfo{author}{R.~P. {Fender}},
\newblock \bibinfo{title}{{A radio jet from the optical and x-ray bright
  stellar tidal disruption flare ASASSN-14li}},
\newblock \bibinfo{journal}{Science} \bibinfo{volume}{351}
  (\bibinfo{year}{2016}) \bibinfo{pages}{62--65}.
\bibitem[{{Fender} et~al.(2004){Fender}, {Belloni}, and
  {Gallo}}]{2004MNRAS.355.1105F}
\bibinfo{author}{R.~P. {Fender}}, \bibinfo{author}{T.~M. {Belloni}},
  \bibinfo{author}{E.~{Gallo}},
\newblock \bibinfo{title}{{Towards a unified model for black hole X-ray binary
  jets}},
\newblock \bibinfo{journal}{MNRAS} \bibinfo{volume}{355} (\bibinfo{year}{2004})
  \bibinfo{pages}{1105--1118}.
\bibitem[{{Tetarenko} et~al.(2017){Tetarenko}, {Sivakoff}, {Miller-Jones},
  {Rosolowsky}, {Petitpas}, {Gurwell}, {Wouterloot}, {Fender}, {Heinz},
  {Maitra}, {Markoff}, {Migliari}, {Rupen}, {Rushton}, {Russell}, {Russell},
  and {Sarazin}}]{2017MNRAS.469.3141T}
\bibinfo{author}{A.~J. {Tetarenko}}, \bibinfo{author}{G.~R. {Sivakoff}},
  \bibinfo{author}{J.~C.~A. {Miller-Jones}}, \bibinfo{author}{E.~W.
  {Rosolowsky}}, \bibinfo{author}{G.~{Petitpas}},
  \bibinfo{author}{M.~{Gurwell}}, \bibinfo{author}{J.~{Wouterloot}},
  \bibinfo{author}{R.~{Fender}}, \bibinfo{author}{S.~{Heinz}},
  \bibinfo{author}{D.~{Maitra}}, \bibinfo{author}{S.~B. {Markoff}},
  \bibinfo{author}{S.~{Migliari}}, \bibinfo{author}{M.~P. {Rupen}},
  \bibinfo{author}{A.~P. {Rushton}}, \bibinfo{author}{D.~M. {Russell}},
  \bibinfo{author}{T.~D. {Russell}}, \bibinfo{author}{C.~L. {Sarazin}},
\newblock \bibinfo{title}{{Extreme jet ejections from the black hole X-ray
  binary V404 Cygni}},
\newblock \bibinfo{journal}{MNRAS} \bibinfo{volume}{469} (\bibinfo{year}{2017})
  \bibinfo{pages}{3141--3162}.
\bibitem[{{Carbone} et~al.(2016){Carbone}, {van der Horst}, {Wijers},
  {Swinbank}, {Rowlinson}, {Broderick}, {Cendes}, {Stewart}, {Bell}, {Breton},
  {Corbel}, {Eislöffel}, {Fender}, {Grie{\ss}meier}, {Hessels}, {Jonker},
  {Kramer}, {Law}, {Miller-Jones}, {Pietka}, {Scheers}, {Stappers}, {van
  Leeuwen}, {Wijnands}, {Wise}, and {Zarka}}]{2016MNRAS.459.3161C}
\bibinfo{author}{D.~{Carbone}}, \bibinfo{author}{A.~J. {van der Horst}},
  \bibinfo{author}{R.~A.~M.~J. {Wijers}}, \bibinfo{author}{J.~D. {Swinbank}},
  \bibinfo{author}{A.~{Rowlinson}}, \bibinfo{author}{J.~W. {Broderick}},
  \bibinfo{author}{Y.~N. {Cendes}}, \bibinfo{author}{A.~J. {Stewart}},
  \bibinfo{author}{M.~E. {Bell}}, \bibinfo{author}{R.~P. {Breton}},
  \bibinfo{author}{S.~{Corbel}}, \bibinfo{author}{J.~{Eislöffel}},
  \bibinfo{author}{R.~P. {Fender}}, \bibinfo{author}{J.~M. {Grie{\ss}meier}},
  \bibinfo{author}{J.~W.~T. {Hessels}}, \bibinfo{author}{P.~{Jonker}},
  \bibinfo{author}{M.~{Kramer}}, \bibinfo{author}{C.~J. {Law}},
  \bibinfo{author}{J.~C.~A. {Miller-Jones}}, \bibinfo{author}{M.~{Pietka}},
  \bibinfo{author}{L.~H.~A. {Scheers}}, \bibinfo{author}{B.~W. {Stappers}},
  \bibinfo{author}{J.~{van Leeuwen}}, \bibinfo{author}{R.~{Wijnands}},
  \bibinfo{author}{M.~{Wise}}, \bibinfo{author}{P.~{Zarka}},
\newblock \bibinfo{title}{{New methods to constrain the radio transient rate:
  results from a survey of four fields with LOFAR}},
\newblock \bibinfo{journal}{MNRAS} \bibinfo{volume}{459} (\bibinfo{year}{2016})
  \bibinfo{pages}{3161--3174}.
\bibitem[{{Carbone} et~al.(2017){Carbone}, {van der Horst}, {Wijers}, and
  {Rowlinson}}]{2017MNRAS.465.4106C}
\bibinfo{author}{D.~{Carbone}}, \bibinfo{author}{A.~J. {van der Horst}},
  \bibinfo{author}{R.~A.~M.~J. {Wijers}}, \bibinfo{author}{A.~{Rowlinson}},
\newblock \bibinfo{title}{{Calculating transient rates from surveys}},
\newblock \bibinfo{journal}{MNRAS} \bibinfo{volume}{465} (\bibinfo{year}{2017})
  \bibinfo{pages}{4106--4117}.
\bibitem[{{Bower} and {Saul}(2011)}]{2011ApJ...728L..14B}
\bibinfo{author}{G.~C. {Bower}}, \bibinfo{author}{D.~{Saul}},
\newblock \bibinfo{title}{{A Search for Radio Transients in Very Large Array
  Archival Images of the 3C 286 Field}},
\newblock \bibinfo{journal}{APJL} \bibinfo{volume}{728} (\bibinfo{year}{2011})
  \bibinfo{pages}{L14}.
\bibitem[{{Astropy Collaboration} et~al.(2013){Astropy Collaboration},
  {Robitaille}, {Tollerud}, {Greenfield}, {Droettboom}, {Bray}, {Aldcroft},
  {Davis}, {Ginsburg}, {Price-Whelan}, {Kerzendorf}, {Conley}, {Crighton},
  {Barbary}, {Muna}, {Ferguson}, {Grollier}, {Parikh}, {Nair}, {Unther},
  {Deil}, {Woillez}, {Conseil}, {Kramer}, {Turner}, {Singer}, {Fox}, {Weaver},
  {Zabalza}, {Edwards}, {Azalee Bostroem}, {Burke}, {Casey}, {Crawford},
  {Dencheva}, {Ely}, {Jenness}, {Labrie}, {Lim}, {Pierfederici}, {Pontzen},
  {Ptak}, {Refsdal}, {Servillat}, and {Streicher}}]{2013A&A...558A..33A}
\bibinfo{author}{{Astropy Collaboration}}, \bibinfo{author}{T.~P.
  {Robitaille}}, \bibinfo{author}{E.~J. {Tollerud}},
  \bibinfo{author}{P.~{Greenfield}}, \bibinfo{author}{M.~{Droettboom}},
  \bibinfo{author}{E.~{Bray}}, \bibinfo{author}{T.~{Aldcroft}},
  \bibinfo{author}{M.~{Davis}}, \bibinfo{author}{A.~{Ginsburg}},
  \bibinfo{author}{A.~M. {Price-Whelan}}, \bibinfo{author}{W.~E. {Kerzendorf}},
  \bibinfo{author}{A.~{Conley}}, \bibinfo{author}{N.~{Crighton}},
  \bibinfo{author}{K.~{Barbary}}, \bibinfo{author}{D.~{Muna}},
  \bibinfo{author}{H.~{Ferguson}}, \bibinfo{author}{F.~{Grollier}},
  \bibinfo{author}{M.~M. {Parikh}}, \bibinfo{author}{P.~H. {Nair}},
  \bibinfo{author}{H.~M. {Unther}}, \bibinfo{author}{C.~{Deil}},
  \bibinfo{author}{J.~{Woillez}}, \bibinfo{author}{S.~{Conseil}},
  \bibinfo{author}{R.~{Kramer}}, \bibinfo{author}{J.~E.~H. {Turner}},
  \bibinfo{author}{L.~{Singer}}, \bibinfo{author}{R.~{Fox}},
  \bibinfo{author}{B.~A. {Weaver}}, \bibinfo{author}{V.~{Zabalza}},
  \bibinfo{author}{Z.~I. {Edwards}}, \bibinfo{author}{K.~{Azalee Bostroem}},
  \bibinfo{author}{D.~J. {Burke}}, \bibinfo{author}{A.~R. {Casey}},
  \bibinfo{author}{S.~M. {Crawford}}, \bibinfo{author}{N.~{Dencheva}},
  \bibinfo{author}{J.~{Ely}}, \bibinfo{author}{T.~{Jenness}},
  \bibinfo{author}{K.~{Labrie}}, \bibinfo{author}{P.~L. {Lim}},
  \bibinfo{author}{F.~{Pierfederici}}, \bibinfo{author}{A.~{Pontzen}},
  \bibinfo{author}{A.~{Ptak}}, \bibinfo{author}{B.~{Refsdal}},
  \bibinfo{author}{M.~{Servillat}}, \bibinfo{author}{O.~{Streicher}},
\newblock \bibinfo{title}{{Astropy: A community Python package for astronomy}},
\newblock \bibinfo{journal}{AAP} \bibinfo{volume}{558} (\bibinfo{year}{2013})
  \bibinfo{pages}{A33}.
\bibitem[{{Virtanen} et~al.(2020){Virtanen}, {Gommers}, {Oliphant},
  {Haberland}, {Reddy}, {Cournapeau}, {Burovski}, {Peterson}, {Weckesser},
  {Bright}, {van der Walt}, {Brett}, {Wilson}, {Millman}, {Mayorov}, {Nelson},
  {Jones}, {Kern}, {Larson}, {Carey}, {Polat}, {Feng}, {Moore}, {VanderPlas},
  {Laxalde}, {Perktold}, {Cimrman}, {Henriksen}, {Quintero}, {Harris},
  {Archibald}, {Ribeiro}, {Pedregosa}, {van Mulbregt}, and {SciPy 1. 0
  Contributors}}]{2020NatMe..17..261V}
\bibinfo{author}{P.~{Virtanen}}, \bibinfo{author}{R.~{Gommers}},
  \bibinfo{author}{T.~E. {Oliphant}}, \bibinfo{author}{M.~{Haberland}},
  \bibinfo{author}{T.~{Reddy}}, \bibinfo{author}{D.~{Cournapeau}},
  \bibinfo{author}{E.~{Burovski}}, \bibinfo{author}{P.~{Peterson}},
  \bibinfo{author}{W.~{Weckesser}}, \bibinfo{author}{J.~{Bright}},
  \bibinfo{author}{S.~J. {van der Walt}}, \bibinfo{author}{M.~{Brett}},
  \bibinfo{author}{J.~{Wilson}}, \bibinfo{author}{K.~J. {Millman}},
  \bibinfo{author}{N.~{Mayorov}}, \bibinfo{author}{A.~R.~J. {Nelson}},
  \bibinfo{author}{E.~{Jones}}, \bibinfo{author}{R.~{Kern}},
  \bibinfo{author}{E.~{Larson}}, \bibinfo{author}{C.~J. {Carey}},
  \bibinfo{author}{{\.I}.~{Polat}}, \bibinfo{author}{Y.~{Feng}},
  \bibinfo{author}{E.~W. {Moore}}, \bibinfo{author}{J.~{VanderPlas}},
  \bibinfo{author}{D.~{Laxalde}}, \bibinfo{author}{J.~{Perktold}},
  \bibinfo{author}{R.~{Cimrman}}, \bibinfo{author}{I.~{Henriksen}},
  \bibinfo{author}{E.~A. {Quintero}}, \bibinfo{author}{C.~R. {Harris}},
  \bibinfo{author}{A.~M. {Archibald}}, \bibinfo{author}{A.~H. {Ribeiro}},
  \bibinfo{author}{F.~{Pedregosa}}, \bibinfo{author}{P.~{van Mulbregt}},
  \bibinfo{author}{{SciPy 1. 0 Contributors}},
\newblock \bibinfo{title}{{SciPy 1.0: fundamental algorithms for scientific
  computing in Python}},
\newblock \bibinfo{journal}{Nature Methods} \bibinfo{volume}{17}
  (\bibinfo{year}{2020}) \bibinfo{pages}{261--272}.
\bibitem[{{McMullin} et~al.(2007){McMullin}, {Waters}, {Schiebel}, {Young}, and
  {Golap}}]{2007ASPC..376..127M}
\bibinfo{author}{J.~P. {McMullin}}, \bibinfo{author}{B.~{Waters}},
  \bibinfo{author}{D.~{Schiebel}}, \bibinfo{author}{W.~{Young}},
  \bibinfo{author}{K.~{Golap}},
\newblock \bibinfo{title}{{CASA Architecture and Applications}},
\newblock in: \bibinfo{editor}{R.~A. {Shaw}}, \bibinfo{editor}{F.~{Hill}},
  \bibinfo{editor}{D.~J. {Bell}} (Eds.), \bibinfo{booktitle}{Astronomical Data
  Analysis Software and Systems XVI}, volume \bibinfo{volume}{376} of
  \textit{\bibinfo{series}{Astronomical Society of the Pacific Conference
  Series}}, \bibinfo{year}{2007}, p. \bibinfo{pages}{127}.
\bibitem[{{Johnson} et~al.(2005){Johnson}, {Kemp}, and
  {Kotz}}]{12005udd3.inbook.....JKK}
\bibinfo{author}{N.~L. {Johnson}}, \bibinfo{author}{A.~W. {Kemp}},
  \bibinfo{author}{S.~{Kotz}}, \bibinfo{title}{{Univariate Discrete
  Distributions}}, \bibinfo{publisher}{Wiley \& Sons, Inc},
  \bibinfo{address}{New York, NY, USA}, \bibinfo{year}{2005}.
\bibitem[{Driessen et~al.(2019)Driessen, McDonald, Buckley, Caleb, Kotze,
  Potter, Rajwade, Rowlinson, Stappers, Tremou, Woudt, Fender, Armstrong,
  Groot, Heywood, Horesh, Horst, Koerding, McBride, Miller-Jones, Mooley, and
  Wijers}]{10.1093/mnras/stz3027}
\bibinfo{author}{L.~N. Driessen}, \bibinfo{author}{I.~McDonald},
  \bibinfo{author}{D.~A.~H. Buckley}, \bibinfo{author}{M.~Caleb},
  \bibinfo{author}{E.~J. Kotze}, \bibinfo{author}{S.~B. Potter},
  \bibinfo{author}{K.~M. Rajwade}, \bibinfo{author}{A.~Rowlinson},
  \bibinfo{author}{B.~W. Stappers}, \bibinfo{author}{E.~Tremou},
  \bibinfo{author}{P.~A. Woudt}, \bibinfo{author}{R.~P. Fender},
  \bibinfo{author}{R.~Armstrong}, \bibinfo{author}{P.~Groot},
  \bibinfo{author}{I.~Heywood}, \bibinfo{author}{A.~Horesh},
  \bibinfo{author}{A.~J. v.~d. Horst}, \bibinfo{author}{E.~Koerding},
  \bibinfo{author}{V.~A. McBride}, \bibinfo{author}{J.~C.~A. Miller-Jones},
  \bibinfo{author}{K.~P. Mooley}, \bibinfo{author}{R.~A. M.~J. Wijers},
\newblock \bibinfo{title}{{MKT J170456.2-482100: the first transient
  discovered by MeerKAT}},
\newblock \bibinfo{journal}{Monthly Notices of the Royal Astronomical Society}
  (\bibinfo{year}{2019}).
\bibitem[{{van Haarlem} et~al.(2013){van Haarlem}, {Wise}, {Gunst}, {Heald},
  {McKean}, {Hessels}, {de Bruyn}, {Nijboer}, {Swinbank}, {Fallows},
  {Brentjens}, {Nelles}, {Beck}, {Falcke}, {Fender}, {Hörandel}, {Koopmans},
  {Mann}, {Miley}, {Röttgering}, {Stappers}, {Wijers}, {Zaroubi}, {van den
  Akker}, {Alexov}, {Anderson}, {Anderson}, {van Ardenne}, {Arts}, {Asgekar},
  {Avruch}, {Batejat}, {B{\"a}hren}, {Bell}, {Bell}, {van Bemmel}, {Bennema},
  {Bentum}, {Bernardi}, {Best}, {B{\^\i}rzan}, {Bonafede}, {Boonstra}, {Braun},
  {Bregman}, {Breitling}, {van de Brink}, {Broderick}, {Broekema}, {Brouw},
  {Brüggen}, {Butcher}, {van Cappellen}, {Ciardi}, {Coenen}, {Conway},
  {Coolen}, {Corstanje}, {Damstra}, {Davies}, {Deller}, {Dettmar}, {van
  Diepen}, {Dijkstra}, {Donker}, {Doorduin}, {Dromer}, {Drost}, {van Duin},
  {Eislöffel}, {van Enst}, {Ferrari}, {Frieswijk}, {Gankema}, {Garrett}, {de
  Gasperin}, {Gerbers}, {de Geus}, {Grie{\ss}meier}, {Grit}, {Gruppen},
  {Hamaker}, {Hassall}, {Hoeft}, {Holties}, {Horneffer}, {van der Horst}, {van
  Houwelingen}, {Huijgen}, {Iacobelli}, {Intema}, {Jackson}, {Jelic}, {de
  Jong}, {Juette}, {Kant}, {Karastergiou}, {Koers}, {Kollen}, {Kondratiev},
  {Kooistra}, {Koopman}, {Koster}, {Kuniyoshi}, {Kramer}, {Kuper},
  {Lambropoulos}, {Law}, {van Leeuwen}, {Lemaitre}, {Loose}, {Maat}, {Macario},
  {Markoff}, {Masters}, {McFadden}, {McKay-Bukowski}, {Meijering}, {Meulman},
  {Mevius}, {Middelberg}, {Millenaar}, {Miller-Jones}, {Mohan}, {Mol},
  {Morawietz}, {Morganti}, {Mulcahy}, {Mulder}, {Munk}, {Nieuwenhuis}, {van
  Nieuwpoort}, {Noordam}, {Norden}, {Noutsos}, {Offringa}, {Olofsson}, {Omar},
  {Orr{\'u}}, {Overeem}, {Paas}, {Pand ey-Pommier}, {Pandey}, {Pizzo},
  {Polatidis}, {Rafferty}, {Rawlings}, {Reich}, {de Reijer}, {Reitsma},
  {Renting}, {Riemers}, {Rol}, {Romein}, {Roosjen}, {Ruiter}, {Scaife}, {van
  der Schaaf}, {Scheers}, {Schellart}, {Schoenmakers}, {Schoonderbeek},
  {Serylak}, {Shulevski}, {Sluman}, {Smirnov}, {Sobey}, {Spreeuw}, {Steinmetz},
  {Sterks}, {Stiepel}, {Stuurwold}, {Tagger}, {Tang}, {Tasse}, {Thomas},
  {Thoudam}, {Toribio}, {van der Tol}, {Usov}, {van Veelen}, {van der Veen},
  {ter Veen}, {Verbiest}, {Vermeulen}, {Vermaas}, {Vocks}, {Vogt}, {de Vos},
  {van der Wal}, {van Weeren}, {Weggemans}, {Weltevrede}, {White}, {Wijnholds},
  {Wilhelmsson}, {Wucknitz}, {Yatawatta}, {Zarka}, {Zensus}, and {van
  Zwieten}}]{2013A&A...556A...2V}
\bibinfo{author}{M.~P. {van Haarlem}}, \bibinfo{author}{M.~W. {Wise}},
  \bibinfo{author}{A.~W. {Gunst}}, \bibinfo{author}{G.~{Heald}},
  \bibinfo{author}{J.~P. {McKean}}, \bibinfo{author}{J.~W.~T. {Hessels}},
  \bibinfo{author}{A.~G. {de Bruyn}}, \bibinfo{author}{R.~{Nijboer}},
  \bibinfo{author}{J.~{Swinbank}}, \bibinfo{author}{R.~{Fallows}},
  \bibinfo{author}{M.~{Brentjens}}, \bibinfo{author}{A.~{Nelles}},
  \bibinfo{author}{R.~{Beck}}, \bibinfo{author}{H.~{Falcke}},
  \bibinfo{author}{R.~{Fender}}, \bibinfo{author}{J.~{Hörandel}},
  \bibinfo{author}{L.~V.~E. {Koopmans}}, \bibinfo{author}{G.~{Mann}},
  \bibinfo{author}{G.~{Miley}}, \bibinfo{author}{H.~{Röttgering}},
  \bibinfo{author}{B.~W. {Stappers}}, \bibinfo{author}{R.~A.~M.~J. {Wijers}},
  \bibinfo{author}{S.~{Zaroubi}}, \bibinfo{author}{M.~{van den Akker}},
  \bibinfo{author}{A.~{Alexov}}, \bibinfo{author}{J.~{Anderson}},
  \bibinfo{author}{K.~{Anderson}}, \bibinfo{author}{A.~{van Ardenne}},
  \bibinfo{author}{M.~{Arts}}, \bibinfo{author}{A.~{Asgekar}},
  \bibinfo{author}{I.~M. {Avruch}}, \bibinfo{author}{F.~{Batejat}},
  \bibinfo{author}{L.~{B{\"a}hren}}, \bibinfo{author}{M.~E. {Bell}},
  \bibinfo{author}{M.~R. {Bell}}, \bibinfo{author}{I.~{van Bemmel}},
  \bibinfo{author}{P.~{Bennema}}, \bibinfo{author}{M.~J. {Bentum}},
  \bibinfo{author}{G.~{Bernardi}}, \bibinfo{author}{P.~{Best}},
  \bibinfo{author}{L.~{B{\^\i}rzan}}, \bibinfo{author}{A.~{Bonafede}},
  \bibinfo{author}{A.~J. {Boonstra}}, \bibinfo{author}{R.~{Braun}},
  \bibinfo{author}{J.~{Bregman}}, \bibinfo{author}{F.~{Breitling}},
  \bibinfo{author}{R.~H. {van de Brink}}, \bibinfo{author}{J.~{Broderick}},
  \bibinfo{author}{P.~C. {Broekema}}, \bibinfo{author}{W.~N. {Brouw}},
  \bibinfo{author}{M.~{Brüggen}}, \bibinfo{author}{H.~R. {Butcher}},
  \bibinfo{author}{W.~{van Cappellen}}, \bibinfo{author}{B.~{Ciardi}},
  \bibinfo{author}{T.~{Coenen}}, \bibinfo{author}{J.~{Conway}},
  \bibinfo{author}{A.~{Coolen}}, \bibinfo{author}{A.~{Corstanje}},
  \bibinfo{author}{S.~{Damstra}}, \bibinfo{author}{O.~{Davies}},
  \bibinfo{author}{A.~T. {Deller}}, \bibinfo{author}{R.~J. {Dettmar}},
  \bibinfo{author}{G.~{van Diepen}}, \bibinfo{author}{K.~{Dijkstra}},
  \bibinfo{author}{P.~{Donker}}, \bibinfo{author}{A.~{Doorduin}},
  \bibinfo{author}{J.~{Dromer}}, \bibinfo{author}{M.~{Drost}},
  \bibinfo{author}{A.~{van Duin}}, \bibinfo{author}{J.~{Eislöffel}},
  \bibinfo{author}{J.~{van Enst}}, \bibinfo{author}{C.~{Ferrari}},
  \bibinfo{author}{W.~{Frieswijk}}, \bibinfo{author}{H.~{Gankema}},
  \bibinfo{author}{M.~A. {Garrett}}, \bibinfo{author}{F.~{de Gasperin}},
  \bibinfo{author}{M.~{Gerbers}}, \bibinfo{author}{E.~{de Geus}},
  \bibinfo{author}{J.~M. {Grie{\ss}meier}}, \bibinfo{author}{T.~{Grit}},
  \bibinfo{author}{P.~{Gruppen}}, \bibinfo{author}{J.~P. {Hamaker}},
  \bibinfo{author}{T.~{Hassall}}, \bibinfo{author}{M.~{Hoeft}},
  \bibinfo{author}{H.~A. {Holties}}, \bibinfo{author}{A.~{Horneffer}},
  \bibinfo{author}{A.~{van der Horst}}, \bibinfo{author}{A.~{van Houwelingen}},
  \bibinfo{author}{A.~{Huijgen}}, \bibinfo{author}{M.~{Iacobelli}},
  \bibinfo{author}{H.~{Intema}}, \bibinfo{author}{N.~{Jackson}},
  \bibinfo{author}{V.~{Jelic}}, \bibinfo{author}{A.~{de Jong}},
  \bibinfo{author}{E.~{Juette}}, \bibinfo{author}{D.~{Kant}},
  \bibinfo{author}{A.~{Karastergiou}}, \bibinfo{author}{A.~{Koers}},
  \bibinfo{author}{H.~{Kollen}}, \bibinfo{author}{V.~I. {Kondratiev}},
  \bibinfo{author}{E.~{Kooistra}}, \bibinfo{author}{Y.~{Koopman}},
  \bibinfo{author}{A.~{Koster}}, \bibinfo{author}{M.~{Kuniyoshi}},
  \bibinfo{author}{M.~{Kramer}}, \bibinfo{author}{G.~{Kuper}},
  \bibinfo{author}{P.~{Lambropoulos}}, \bibinfo{author}{C.~{Law}},
  \bibinfo{author}{J.~{van Leeuwen}}, \bibinfo{author}{J.~{Lemaitre}},
  \bibinfo{author}{M.~{Loose}}, \bibinfo{author}{P.~{Maat}},
  \bibinfo{author}{G.~{Macario}}, \bibinfo{author}{S.~{Markoff}},
  \bibinfo{author}{J.~{Masters}}, \bibinfo{author}{R.~A. {McFadden}},
  \bibinfo{author}{D.~{McKay-Bukowski}}, \bibinfo{author}{H.~{Meijering}},
  \bibinfo{author}{H.~{Meulman}}, \bibinfo{author}{M.~{Mevius}},
  \bibinfo{author}{E.~{Middelberg}}, \bibinfo{author}{R.~{Millenaar}},
  \bibinfo{author}{J.~C.~A. {Miller-Jones}}, \bibinfo{author}{R.~N. {Mohan}},
  \bibinfo{author}{J.~D. {Mol}}, \bibinfo{author}{J.~{Morawietz}},
  \bibinfo{author}{R.~{Morganti}}, \bibinfo{author}{D.~D. {Mulcahy}},
  \bibinfo{author}{E.~{Mulder}}, \bibinfo{author}{H.~{Munk}},
  \bibinfo{author}{L.~{Nieuwenhuis}}, \bibinfo{author}{R.~{van Nieuwpoort}},
  \bibinfo{author}{J.~E. {Noordam}}, \bibinfo{author}{M.~{Norden}},
  \bibinfo{author}{A.~{Noutsos}}, \bibinfo{author}{A.~R. {Offringa}},
  \bibinfo{author}{H.~{Olofsson}}, \bibinfo{author}{A.~{Omar}},
  \bibinfo{author}{E.~{Orr{\'u}}}, \bibinfo{author}{R.~{Overeem}},
  \bibinfo{author}{H.~{Paas}}, \bibinfo{author}{M.~{Pand ey-Pommier}},
  \bibinfo{author}{V.~N. {Pandey}}, \bibinfo{author}{R.~{Pizzo}},
  \bibinfo{author}{A.~{Polatidis}}, \bibinfo{author}{D.~{Rafferty}},
  \bibinfo{author}{S.~{Rawlings}}, \bibinfo{author}{W.~{Reich}},
  \bibinfo{author}{J.~P. {de Reijer}}, \bibinfo{author}{J.~{Reitsma}},
  \bibinfo{author}{G.~A. {Renting}}, \bibinfo{author}{P.~{Riemers}},
  \bibinfo{author}{E.~{Rol}}, \bibinfo{author}{J.~W. {Romein}},
  \bibinfo{author}{J.~{Roosjen}}, \bibinfo{author}{M.~{Ruiter}},
  \bibinfo{author}{A.~{Scaife}}, \bibinfo{author}{K.~{van der Schaaf}},
  \bibinfo{author}{B.~{Scheers}}, \bibinfo{author}{P.~{Schellart}},
  \bibinfo{author}{A.~{Schoenmakers}}, \bibinfo{author}{G.~{Schoonderbeek}},
  \bibinfo{author}{M.~{Serylak}}, \bibinfo{author}{A.~{Shulevski}},
  \bibinfo{author}{J.~{Sluman}}, \bibinfo{author}{O.~{Smirnov}},
  \bibinfo{author}{C.~{Sobey}}, \bibinfo{author}{H.~{Spreeuw}},
  \bibinfo{author}{M.~{Steinmetz}}, \bibinfo{author}{C.~G.~M. {Sterks}},
  \bibinfo{author}{H.~J. {Stiepel}}, \bibinfo{author}{K.~{Stuurwold}},
  \bibinfo{author}{M.~{Tagger}}, \bibinfo{author}{Y.~{Tang}},
  \bibinfo{author}{C.~{Tasse}}, \bibinfo{author}{I.~{Thomas}},
  \bibinfo{author}{S.~{Thoudam}}, \bibinfo{author}{M.~C. {Toribio}},
  \bibinfo{author}{B.~{van der Tol}}, \bibinfo{author}{O.~{Usov}},
  \bibinfo{author}{M.~{van Veelen}}, \bibinfo{author}{A.~J. {van der Veen}},
  \bibinfo{author}{S.~{ter Veen}}, \bibinfo{author}{J.~P.~W. {Verbiest}},
  \bibinfo{author}{R.~{Vermeulen}}, \bibinfo{author}{N.~{Vermaas}},
  \bibinfo{author}{C.~{Vocks}}, \bibinfo{author}{C.~{Vogt}},
  \bibinfo{author}{M.~{de Vos}}, \bibinfo{author}{E.~{van der Wal}},
  \bibinfo{author}{R.~{van Weeren}}, \bibinfo{author}{H.~{Weggemans}},
  \bibinfo{author}{P.~{Weltevrede}}, \bibinfo{author}{S.~{White}},
  \bibinfo{author}{S.~J. {Wijnholds}}, \bibinfo{author}{T.~{Wilhelmsson}},
  \bibinfo{author}{O.~{Wucknitz}}, \bibinfo{author}{S.~{Yatawatta}},
  \bibinfo{author}{P.~{Zarka}}, \bibinfo{author}{A.~{Zensus}},
  \bibinfo{author}{J.~{van Zwieten}},
\newblock \bibinfo{title}{{LOFAR: The LOw-Frequency ARray}},
\newblock \bibinfo{journal}{AAP} \bibinfo{volume}{556} (\bibinfo{year}{2013})
  \bibinfo{pages}{A2}.
\bibitem[{{Dewdney} et~al.(2009){Dewdney}, {Hall}, {Schilizzi}, and
  {Lazio}}]{2009IEEEP..97.1482D}
\bibinfo{author}{P.~E. {Dewdney}}, \bibinfo{author}{P.~J. {Hall}},
  \bibinfo{author}{R.~T. {Schilizzi}}, \bibinfo{author}{T.~J.~L.~W. {Lazio}},
\newblock \bibinfo{title}{{The Square Kilometre Array}},
\newblock \bibinfo{journal}{IEEE Proceedings} \bibinfo{volume}{97}
  (\bibinfo{year}{2009}) \bibinfo{pages}{1482--1496}.
\bibitem[{{Murphy} et~al.(2018){Murphy}, {Bolatto}, {Chatterjee}, {Casey},
  {Chomiuk}, {Dale}, {de Pater}, {Dickinson}, {Francesco}, {Hallinan},
  {Isella}, {Kohno}, {Kulkarni}, {Lang}, {Lazio}, {Leroy}, {Loinard},
  {Maccarone}, {Matthews}, {Osten}, {Reid}, {Riechers}, {Sakai}, {Walter}, and
  {Wilner}}]{2018ASPC..517....3M}
\bibinfo{author}{E.~J. {Murphy}}, \bibinfo{author}{A.~{Bolatto}},
  \bibinfo{author}{S.~{Chatterjee}}, \bibinfo{author}{C.~M. {Casey}},
  \bibinfo{author}{L.~{Chomiuk}}, \bibinfo{author}{D.~{Dale}},
  \bibinfo{author}{I.~{de Pater}}, \bibinfo{author}{M.~{Dickinson}},
  \bibinfo{author}{J.~D. {Francesco}}, \bibinfo{author}{G.~{Hallinan}},
  \bibinfo{author}{A.~{Isella}}, \bibinfo{author}{K.~{Kohno}},
  \bibinfo{author}{S.~R. {Kulkarni}}, \bibinfo{author}{C.~{Lang}},
  \bibinfo{author}{T.~J.~W. {Lazio}}, \bibinfo{author}{A.~K. {Leroy}},
  \bibinfo{author}{L.~{Loinard}}, \bibinfo{author}{T.~J. {Maccarone}},
  \bibinfo{author}{B.~C. {Matthews}}, \bibinfo{author}{R.~A. {Osten}},
  \bibinfo{author}{M.~J. {Reid}}, \bibinfo{author}{D.~{Riechers}},
  \bibinfo{author}{N.~{Sakai}}, \bibinfo{author}{F.~{Walter}},
  \bibinfo{author}{D.~{Wilner}},
\newblock \bibinfo{title}{{The ngVLA Science Case and Associated Science
  Requirements}},
\newblock in: \bibinfo{editor}{E.~{Murphy}} (Ed.), \bibinfo{booktitle}{Science
  with a Next Generation Very Large Array}, volume \bibinfo{volume}{517} of
  \textit{\bibinfo{series}{Astronomical Society of the Pacific Conference
  Series}}, \bibinfo{year}{2018}, p.~\bibinfo{pages}{3}.

\end{thebibliography}





\end{document}